\documentclass[twocolumn,pre,superscriptaddress,10pt,aps,floatfix]{revtex4-1}

\usepackage[nice]{nicefrac}
\usepackage{graphicx,color}
\usepackage{amsmath,amssymb,bm,bbm}
\usepackage{amsmath}
\usepackage{braket}
\usepackage{bbold}
\usepackage{float}
\usepackage{placeins}
\usepackage{soul}
\usepackage{enumitem}
\usepackage{mathtools}
\usepackage{todonotes}
\usepackage{subdepth}

\usepackage[plainpages=false,pdfpagelabels,colorlinks=true,linkcolor=red,urlcolor=blue,citecolor=blue,pdftitle={},pdfauthor={},pdfdisplaydoctitle=true,pdfduplex=DuplexFlipLongEdge]{hyperref}

\definecolor{darkred}{rgb}{0.90,0.2,0.2}
\definecolor{darkgreen}{rgb}{0,0.60,.2}
\definecolor{darkblue}{rgb}{0.1,0.3,1}
\definecolor{grey}{cmyk}{0,0,0,0.25}
\definecolor{orange}{cmyk}{0,0.6,0.8,0}

\begin{document}

\title{Critical quantum dynamics of observables at eigenstate transitions
}

\author{Simon Jiricek}
\affiliation{Institut f\"{u}r Theoretische Physik, Georg-August-Universit\"at G\"ottingen, D-37077 G\"ottingen, Germany}
\author{Miroslav Hopjan}
\affiliation{Department of Theoretical Physics, J. Stefan Institute, SI-1000 Ljubljana, Slovenia}
\author{Patrycja  \L yd\.{z}ba}
\affiliation{Department of Theoretical Physics, Wroclaw University of Science and Technology, 50-370 Wroc{\l}aw, Poland}
\author{Fabian Heidrich-Meisner}
\affiliation{Institut f\"{u}r Theoretische Physik, Georg-August-Universit\"at G\"ottingen, D-37077 G\"ottingen, Germany}
\author{Lev Vidmar}
\affiliation{Department of Theoretical Physics, J. Stefan Institute, SI-1000 Ljubljana, Slovenia}
\affiliation{Department of Physics, Faculty of Mathematics and Physics, University of Ljubljana, SI-1000 Ljubljana, Slovenia\looseness=-1}

\begin{abstract}
It is an outstanding goal to unveil the key features of quantum dynamics at eigenstate transitions. 
Focusing on quadratic fermionic Hamiltonians that exhibit localization transitions, we identify physical observables that exhibit scale-invariant critical dynamics at the transition when quenched from the initially localized charge density-wave states.
The identification is based on two ingredients:
(a) a relationship between the time evolution of observables in a many-body state and the transition probabilities of single-particle states, and (b)  scale invariance of transition probabilities, which generalizes the corresponding recent result for survival probabilities [M. Hopjan and L. Vidmar, \href{https://doi.org/10.1103/PhysRevLett.131.060404}{Phys.~Rev.~Lett.~\textbf{131}, 060404  (2023)}; \href{https://doi.org/10.1103/PhysRevResearch.5.043301}{Phys.~Rev.~Res.~\textbf{5}, 043301  (2023)}].
These properties suggest that the scale-invariant critical dynamics in the quantum-quench dynamics is also exhibited by the observables, which share the common eigenbasis with the Hamiltonian before the quench.
Focusing on experimentally relevant observables such as site occupations and the particle imbalance, we numerically demonstrate their critical behavior at the eigenstate transitions in the three-dimensional Anderson model and the one-dimensional Aubry–André model.
\end{abstract}
\maketitle

\section{Introduction}

Research in recent decades contributed significantly to our understanding of how an interacting quantum system, evolving in isolation from any environment, approaches a thermal equilibrium~\cite{deutsch_91, srednicki_94, rigol_dunjko_08, dalessio_kafri_16, mori_ikeda_18}.
Experimentally, this can be conveniently tested via a so-called quantum quench, in which a far-from-equilibrium initial state evolves unitarily under a time-independent Hamiltonian~\cite{greiner02, kinoshita_wenger_06, polkovnikov_sengupta_11, eisert_friesdorf_15}.
Then, thermalization of a system can be monitored either via the expectation values of observables approaching predictions of the Gibbs ensembles~\cite{Trotzky2012, clos_porras_16, tang_kao_18} or by generating maximal entanglement between different parts of the systems, given the global constraints such as the total energy~\cite{Kaufman2016, Neill2016}.

Detecting  the boundaries of thermalizing behavior, at which ergodicity-breaking phase transitions may take place~\cite{Schreiber15, smith_lee_16, rispoli_lukin_19, Guo20, gong_moraesneto_21, chiaro_neill_22, leonard_kim_23}, turns out to be even more challenging.
On the theory side, such transitions are most commonly studied via the properties of Hamiltonian eigenstates and spectra~\cite{oganesyan_huse_07, pal_huse_10}, which exhibit a sharp change at the transition and, therefore, these transitions are dubbed {\it eigenstate transitions}.
From the dynamical perspective, a lesson learned from Anderson localization~\cite{anderson_58} is the following:
in a time-dependent problem, localization can only unambiguously be established at the longest physically relevant relaxation time that exhibits the same scaling with system size as the corresponding Heisenberg time, i.e., the inverse level spacing~\cite{edwards_thouless_72, sierant_delande_20, suntajs_prosen_21}.
In this way, features incompatible with localization, e.g., diffusive transport, can be ruled out.
In interacting systems, the Heisenberg time related to this criterion is very long since it increases exponentially with the number of lattice sites~\cite{suntajs_bonca_20a}.
A puzzling question then appears: Does one really need to wait for such a long time to determine whether a given system is localized or ergodic?

These arguments motivate the quest for establishing alternative ways to detect critical behavior in a system's nonequilibrium dynamics, at times shorter than the Heisenberg time. This possibility has  recently been opened by the observation, at the critical point, of scale-invariant properties of long-range spectral correlations measured by the spectral form factor~\cite{suntajs_prosen_21,suntajs_vidmar_22, hopjan2023, hopjan2023scaleinvariant} and survival probabilities~\cite{hopjan2023, hopjan2023scaleinvariant}.
Remarkably, these features emerge in the so-called {\it mid-time} dynamics~\cite{hopjan2023scaleinvariant}, i.e., at times much shorter than
the Heisenberg time.
Even though advanced experimental protocols have been proposed to measure both the spectral form factor~\cite{Vasilyev20, Joshi22} and the survival probability~\cite{LozanoNegro2021,das2024proposal}, an urgent goal of current research is to establish a framework to detect criticality in the mid-time dynamics of {\it few-body} observables.
A particular observable relevant for this work is the particle imbalance on even versus odd lattice sites, which is routinely measured in state-of-the-art quantum-simulator experiments~\cite{Schreiber15,Choi16,Luschen17,Bordia17,Kohlert19,Rubio-Abadal19,Guo20}.

In this paper, we take a step forward in relating the critical dynamics of conventional eigenstate-transition indicators, such as the spectral form factor~\cite{suntajs_prosen_21,suntajs_vidmar_22, hopjan2023, hopjan2023scaleinvariant, berry_85, kos_ljubotina_18, chan_deluca_18a, gharibyan_hanada_18, bertini_kos_18, Liao20, Winer20} and survival probabilities~\cite{hopjan2023, hopjan2023scaleinvariant, Ketzmerick_92, Huckestein94, Schofield_95, Schofield_96, Brandes96, Ketzmerick_97, Ohtsuki97, Gruebele_98, Ng_06, torresherrera_santos_14, torresherrera_santos_15, Leitner_15, Santos2017, torresherrera_garciagarcia_18, Bera2018, Prelovsek18, Schiulaz_19, Karmakar_19, Lezama_21}, with one-body observables.
{Achieving this goal in interacting models exhibiting an ergodicity-breaking phase transition is a challenging task that is beyond the scope of this work. Here we focus on quadratic fermionic Hamiltonians.}
Specifically, we identify a class of observables that exhibit  scale-invariant mid-time dynamics at criticality. These observables share a common eigenbasis with the Hamiltonian before the quantum quench and, as a consequence, their dynamics in many-body states is related to the transition probabilities of initially occupied single-particle states.
Examples of such initial states, which are experimentally accessible and also the focus of this work, are those that form a charge density-wave pattern~\cite{Schreiber15}. We then study two paradigmatic quadratic models that exhibit eigenstate transitions: the three-dimensional (3D) Anderson model and the one-dimensional (1D) Aubry–André model, which both exhibit a transition to localized single-particle states in the site occupation basis at sufficiently large random or quasiperiodic disorder, respectively.

The dynamics of transition probabilities generalize the concept of survival probabilities, for which the notion of scale-invariant mid-time dynamics at criticality has recently been established~\cite{hopjan2023, hopjan2023scaleinvariant}.
Studying the quench dynamics from the initial site-localized charge-density-wave states, we numerically observe scale invariance for the transition probabilities, which shares certain features with the scale-invariant survival probabilities, such as a power-law decay with a similar decay exponent.
The emergence of scale invariance in the mid-time dynamics of transition probabilities appears to be related to the emergence of scale invariance in the dynamics of observables.

The above considerations provide the basis for the main result of this paper: the observation of scale-invariant dynamics of one-body observables, such as site occupations and particle imbalance, in both models at the critical point.
Remarkably, the scale invariance turns out to be a property of the critical point and is not visible away from it. Therefore, it establishes a route to detect the critical point at comparatively short times after a quantum quench.
Our results also lead to the question whether the exponent of the temporal power-law decay of observables is related to the fractal dimension of the initial single-particle states in the Hamiltonian eigenbasis.
We find numerical indications that this may indeed be the case for the 3D Anderson model.

The paper is organized as follows.
In Sec.~\ref{sec:models}, we introduce the models under investigation, and in Sec.~\ref{sec:relationship}, we establish the formal relationship between the quench dynamics of observables in many-body states and the dynamics of transition probabilities in single-particle states.
In Sec.~\ref{sec:case_study}, we discuss a case study of particular initial state that forms a charge density-wave pattern. The scale invariance of transition probabilities is numerically demonstrated in Sec.~\ref{sec:transition_prob} and the connection to the dynamics of survival probabilities is discussed.
The main results are then presented in Sec.~\ref{sec:observables}, in where we establish the existence of  
scale-invariant mid-time dynamics of the particle imbalance at the critical point of the models under investigation.
We conclude in Sec.~\ref{sec:conclusions}.


\section{Models} \label{sec:models}

We study two quadratic models of spinless fermions with particle number conservation that exhibit transitions to real-space localization, given by the Hamiltonian
\begin{equation}
\label{eq:hamnonint}
\hat H= -{J}\sum_{\langle ij\rangle}^{} (\hat{c}_{i}^{\dagger}\hat{c}_{j}^{}+ \hat c_j^\dagger \hat c_i) + \sum_{i=1}^{D}\epsilon_{i}\hat{n}_{i}^{}\;,
\end{equation}
where $\hat{c}_{i}^{\dagger}$ ($\hat{c}_{i}^{}$) are the fermionic creation (annihilation) operators at site $i$, $\langle ij\rangle$ runs over nearest neighbors, $J$ is the hopping matrix element, $\epsilon_{i}$ is the on-site energy and $\hat{n}_{i}^{}=\hat{c}_{i}^{\dagger}\hat{c}_{i}^{}$ is the site occupation operator.

The first model that we study is the Anderson model~\cite{anderson_58, Abrahams79,evers_mirlin_08} on a three-dimensional (3D) cubic lattice with linear size $L$ and the single-particle Hilbert space dimension $D=L^3$. In this model, the on-site energies are drawn from a box distribution $\epsilon_i \in [-W/2,W/2]$ and are independently and identically distributed. We consider periodic boundary conditions.
We note that our numerical results for the 3D Anderson model are averaged over $N_{\rm R}$ different Hamiltonian realizations.
If $N_{\rm R}$ is not explicitly stated in a figure caption, then the corresponding values are those provided in Appendix~\ref{sec:disorder_statistics_3D_Anderson}.

Properties of the critical point of the 3D Anderson model were studied from various angles~\cite{kramer_mackinnon_93, MacKinnon81, MacKinnon83, suntajs_prosen_21, Tarquini2017}, with the
consensus that the system is localized, i.e., insulating for $W >W_{c} \approx 16.5\,J$~\cite{Ohtsuki18}, while below $W_{c}$ it becomes diffusive~\cite{Ohtsuki97,Zhao20,Herbrych21}.
The model exhibits subdiffusion~\cite{Ohtsuki97} and multifractal single-particle  eigenfunctions~\cite{evers_mirlin_08, Rodriguez09, Rodriguez10} at the critical point.
The latter is energy dependent, i.e., at $W >W_{c}$ all single-particle states are localized, at $W=W_{c}$ the
transition occurs in the middle of the band, and at $W <W_{c}$ the system exhibits  mobility edges that are pushed towards the edges of the band with  decreasing disorder~\cite{Schubert_05}.
The single-particle eigenstate transition is also reflected in many-body setups, e.g., in entanglement properties of many-body eigenstates~\cite{Li16},  the dynamics of the entanglement~\cite{Zhao20}, the distribution of the local occupations~\cite{hopjan_orso_21}, and the quantum surface roughness~\cite{bhakuni2023dynamic}.

The second model is the Aubry–André model on a one-dimensional (1D) lattice with $L$ sites and the single-particle Hilbert space dimension $D=L$.
It is subject to the quasiperiodic on-site potential $\epsilon_{i}=\lambda \cos(2\pi q i+\phi)$, where $\lambda$ stands for the amplitude of the potential, $\phi$ is a global phase, and $q=\frac{\sqrt{5}-1}{2}$ is chosen to be the golden ratio. 
All numerical results for the 1D Aubry–André model are averaged over 50 realizations with different values of the global phase $\phi$.

The 1D Aubry–André model exhibits a sharp self-dual delocalization-localization transition at $\lambda_c = 2J$~\cite{Aubry80, Suslov82, Kohmoto83, Chao86, Kohmoto87, Siebesma1987, Hiramoto89, Hiramoto92, Macia2014, Li16, wu2021}. At $\lambda>\lambda_c$ all states are localized in real space, at $\lambda=\lambda_c$ both the eigenspectrum and eigenstates are multifractal, and at $\lambda<\lambda_c$ all states are delocalized. At the critical point, the model is related to the Harper-Hofstadter model describing the motion of an electron in an isotropic two-dimensional lattice under a magnetic field$~$\cite{Harper_1955} giving it a certain two-dimensional character, such as signatures of diffusion$~$\cite{Geisel_91} or the $\propto L^2$ scaling of the typical Heisenberg time$~$\cite{hopjan2023}. This transition was also observed experimentally using cold atoms~\cite{Roati08,Luschen18} and photonic lattices~\cite{Lahini:PRL2009}.
The single-particle eigenstate transition is also reflected in many-body setups, e.g., in entanglement properties of many-body eigenstates~\cite{Li16}, the fractal dimensions of many-body eigenstates~\cite{DeTomasi21}, the dynamics of the entanglement Hamiltonian~\cite{Roy2021a,Roy2021b}, or the quantum surface roughness~\cite{aditya2023familyvicsek} {(see also~\cite{bhakuni2023dynamic})}.


\section{Observables and transition probabilities} \label{sec:relationship}

\subsection{Time evolution of one-body observables}

We are interested in the quantum-quench protocol, in which the system is prepared in a many-body eigenstate $|\Psi_0\rangle$ of the initial quadratic Hamiltonian $\hat{H}_0$ with $N$ particles, and then time evolved under the final quadratic Hamiltonian $\hat{H} = \sum_{q=1}^D \varepsilon_q \hat f_q^\dagger \hat f_q$, where $\hat{f}_{q}^{\dagger}$ ($\hat{f}_{q}^{}$) are the fermionic creation (annihilation) operators of a particle in a single-particle energy eigenstate $|q\rangle$ with eigenvalue $\varepsilon_q$.

As a first step, we express a one-body observable $\hat o$ in its single-particle eigenbasis.
Observables such as the occupation of a single lattice site or a single quasimomentum mode have only a single nonzero eigenvalue.
We refer to this type of observables as rank ${\bf O}(1)$ one-body observables~\cite{Lydzba23}.
They are the focus of this section (Sec.~\ref{sec:relationship}) and we express them as
\begin{equation}
\hat o = \hat n_\rho = \hat c_\rho^\dagger \hat c_\rho\;, 
\end{equation}
where $\hat{c}_\rho^{\dagger}$ ($\hat{c}_\rho$) creates (annihilates) a particle in a single-particle state $\ket{\rho}$. 
In the single-particle Hilbert space, one can also express $\hat{n}_\rho$ as $\hat{n}_\rho\equiv\ket{\rho}\bra{\rho}$. We emphasize that the rank ${\bf O}(1)$ one-body observables are, more generally, those observables that have ${\bf O}(1)$ non-degenerate eigenvalues in their single-particle eigenbasis. Simultaneously, other one-body observables have rank ${\bf O}(D)$ and, hence, can be expressed as ${\bf O}(D)$ sums of rank ${\bf O}(1)$ one-body observables, i.e., $\hat O = \sum_\rho a_\rho \hat n_\rho$.

After a unitary transformation, one can express $\hat o$ in the Hamiltonian single-particle eigenbasis as
\begin{equation} \label{def_o_matele}
    \hat o = \sum_{q\prime, q = 1}^D o_{q\prime q} \hat f_{q\prime}^\dagger \hat f_q \;,
\end{equation}
where $ o_{q\prime q} = \langle q\prime| \hat o |q\rangle$ are the matrix elements in the single-particle Hamiltonian eigenbasis.
The time evolution of the expectation value of $\hat{o}$ in the quantum-quench protocol is given by (setting $\hbar \equiv 1$)
\begin{equation} \label{def_Ot}
    o(t) = \langle \Psi_0| e^{i\hat H t} \hat o e^{-i\hat H t} |\Psi_0\rangle = \sum_{q,q\prime =1}^D o_{q\prime q} R_{q\prime q} e^{-it(\varepsilon_{q\prime}-\varepsilon_q)} \;,
\end{equation}
where the information about the many-body initial state is encoded in the one-body correlation matrix~\cite{Venuti_2013}
\begin{equation} \label{def_Rqq}
    R_{q\prime q} = \langle \Psi_0 | \hat f_{q\prime}^\dagger \hat f_q |\Psi_0\rangle \;.
\end{equation}
One can show that Eqs.~(\ref{def_Ot}) and~(\ref{def_Rqq}) imply the long-time average of $o(t)$ to coincide with the predictions of the generalized Gibbs ensemble~\cite{rigol_dunjko_07, vidmar16, essler_fagotti_16, Ziraldo_2012, He_2013, Lydzba23}.

\subsection{Single-particle transition probabilities}

We now focus on the transition probabilities of an initial single-particle state $|\alpha_B\rangle = \hat c_{\alpha_B}^\dagger |\emptyset\rangle$ to $|\alpha_A\rangle$.
In this subsection, the single-particle states $|\alpha_A\rangle$ and $|\alpha_B\rangle$ do not need to coincide with the eigenstates of the Hamiltonian or observable of interest.
We define the transition probabilities as
\begin{equation} \label{def_Pab}
    P_{AB}(t) = |\langle\alpha_A | e^{-i\hat Ht}|\alpha_B\rangle|^2 \;.
\end{equation}
If $A=B$, $P_{AB}(t)$ from Eq.~(\ref{def_Pab}) represents the survival probability studied in Refs.~\cite{Ketzmerick_92, Huckestein94, Schofield_95, Schofield_96, Brandes96, Ketzmerick_97, Ohtsuki97, Gruebele_98, Ng_06, torresherrera_santos_14, torresherrera_santos_15, Leitner_15, Santos2017, torresherrera_garciagarcia_18, Bera2018, Prelovsek18, Schiulaz_19, Karmakar_19,Lezama_21, hopjan2023, hopjan2023scaleinvariant},
while here we are generally interested in the case $A\neq B$
{~\cite{Torres-Herrera17}}.
Due to the single-particle nature of our setup, one can express $P_{AB}(t)$ as
\begin{equation} \label{def_Pab_long}
    P_{AB}(t) = \sum_{q,q\prime =1}^D o_{qq\prime}^{\alpha_A} o_{q\prime q}^{\alpha_B} e^{-it(\varepsilon_{q\prime}-\varepsilon_q)} \;,
\end{equation}
where $o_{qq\prime}^{\alpha_A} = \langle q |\hat n_{\alpha_A}|q\prime \rangle$ are the matrix elements of the one-body operator $\hat n_{\alpha_A} = \hat c_{\alpha_A}^\dagger \hat c_{\alpha_A}$ in the single-particle Hamiltonian eigenstates.

As a side remark, we note that the transition probability $P_{AB}(t)$ from Eq.~(\ref{def_Pab}) can also be expressed via the two-point correlation function,
\begin{equation}
    P_{AB}(t) = \langle \hat{\cal N}_A(t) \hat{\cal N}_B \rangle = \frac{1}{D} {\rm Tr}\{\hat{\cal N}_A(t) \hat{\cal N}_B \} \;,
\end{equation}
where the trace is carried out in the single-particle Hilbert space and $\hat{\cal N}_A = \sqrt{D} \hat n_{\alpha_A}$.
The role of the rescaling with $\sqrt{D}$ in the latter is to invoke the unit Hilbert-Schmidt norm of the observable, such that it satisfies the single-particle eigenstate thermalization~\cite{lydzba_zhang_21, ulcakar_vidmar_22, Tokarczyk_2023}.

\subsection{From the dynamics of observables to transition probabilities}

The time evolution of a one-body observable $o(t)$ in many-body eigenstates [see Eq.~(\ref{def_Ot})] and the transition probabilities $P_{AB}(t)$ of single-particle states [see Eq.~(\ref{def_Pab_long})], share some similarities. These similarities are particularly apparent when the observable's eigenbasis coincides with the eigenbasis of the initial Hamiltonian $\hat H_0$.
In this case, one can relate the observable $\hat o$ to the operator $\hat n_{\alpha_B}$, 
and hence $o_{q\prime q}$ from Eq.~(\ref{def_Ot}) equals $o_{q\prime q}^{\alpha_B}$ Eq.~(\ref{def_Pab_long}).
It then remains to be determined what the relationship between $R_{q\prime q}$ from Eq.~(\ref{def_Ot}) and $o_{qq\prime}^{\alpha_A}$ from Eq.~(\ref{def_Pab_long}) is.

We focus on the initial many-body state $|\Psi_0\rangle$ in the quantum-quench protocol that is a product of $N$ occupied single-particle states
\begin{equation}
    |\Psi_0\rangle = \prod_{\alpha_l \in \underline{\alpha}} \hat c_{\alpha_l}^\dagger |\emptyset\rangle \;,
\end{equation}
where $\underline{\alpha}$ denotes the set of occupied single-particle states. If one rewrites the operators $\hat f_q^\dagger$, $\hat f_q$ as
\begin{equation}
    \hat f_{q\prime}^\dagger \hat f_q = \sum_{\alpha\prime,\alpha=1}^D o_{qq\prime}^{\alpha\alpha\prime} \hat c_{\alpha\prime}^\dagger \hat c_\alpha \;,
    \;\;\; o_{qq\prime}^{\alpha\alpha\prime} = \langle q | \hat c_\alpha^\dagger \hat c_{\alpha\prime} |q\prime\rangle \;,
\end{equation}
and inserts them into the matrix elements $R_{q\prime q}$ from Eq.~(\ref{def_Rqq}), it becomes clear that $R_{q\prime q}$ can be expressed as
\begin{equation} \label{def_R_qqprime_new}
    R_{q\prime q} = \sum_{\alpha\prime \in \underline{\alpha}} o_{qq\prime}^{\alpha\prime} \;,
\end{equation}
where $o_{qq\prime}^{\alpha\prime} = \delta_{\alpha\alpha\prime} o_{qq\prime}^{\alpha\alpha\prime}$. Inserting Eq.~(\ref{def_R_qqprime_new}) into Eq.~(\ref{def_Ot}), one can relate the time evolution of an observable $\hat o_{A} = \hat c_{\alpha_A}^\dagger \hat c_{\alpha_A}$ in a many-body state to the single-particle transition probabilities,
\begin{equation} \label{def_oA_short}
    o_A(t) = \sum_{\alpha_B \in \underline{\alpha}}  P_{AB}(t)\;.
\end{equation}
When the initial single-particle state $A$ is occupied, one may decompose this expression as
\begin{equation} \label{def_oA_final}
    o_A(t) = P_{AA}(t) + \sum_{\substack{\alpha_B \in \underline{\alpha} \\ A\neq B}}^{}  P_{AB}(t)\;,
\end{equation}
where the first term represents the single-particle survival probability and the second term represents the single-particle transition probabilities.

\begin{figure}[!t]
\includegraphics[width=0.98\columnwidth]{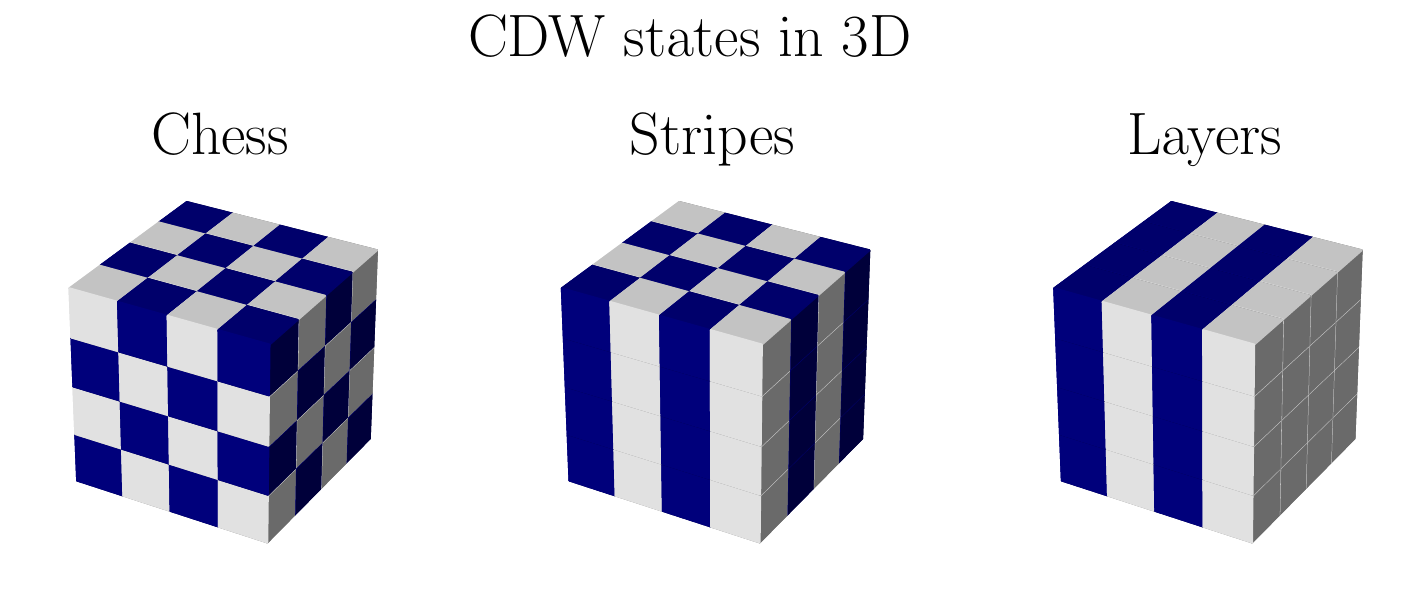}
\vspace{-0.2cm}
\caption{{Sketch of the initial states for the 3D Anderson model ($L=4$) considered in this study. The "chess", "stripes" and "layers" states have the charge-density wave (CDW) pattern of densities in three, two and one directions, respectively. The ``chess'' states are considered in the case study of Sec.~\ref{sec:case_study} and, unless explicitly mentioned otherwise, in the numerical results of all subsequent figures.}
}
\label{sketch}
\end{figure}

Equation~(\ref{def_oA_final}) is the main result of this section and will be instrumental for the subsequent analysis of the critical quantum dynamics of one-body observables.
It connects the time evolution of a one-body observable $\hat o_A$ in a many-body state $|\Psi(t)\rangle$ with the sum of single-particle contributions, each of them describing the transition probabilities from the state $|\alpha_B\rangle$ to the state $|\alpha_A\rangle$.

\section{Case study: Occupations in charge-density wave states}\label{sec:case_study}

{The purpose of this section to illustrate the concepts, which were introduced in the previous section, for a particular initial state, and to motivate a separate study of the transition probabilities in the next section.}
To this end, we set the one-body observable $\hat o_A$ to be the occupation $\hat n_j$ of a lattice site $j$, i.e., $\hat c_{\alpha_A}^\dagger \hat c_{\alpha_A} \to \hat c_j^\dagger \hat c_j$.
We consider the time evolution with the Hamiltonian $\hat H$ from Eq.~\eqref{eq:hamnonint}, quenched from the initial Hamiltonian $\hat H_0= \sum_{i=1}^{D}\epsilon_{i}\hat{n}_{i}^{}$. The initial Hamiltonian $\hat H_0$ corresponds to the second term of $\hat H$ and can be identified as the infinite-disorder limit of $\hat H$. Its single-particle eigenstates are expressed as $|j\rangle = \hat c_j^\dagger |\emptyset\rangle$.
The dynamics is initiated in a prototypical charge-density wave (CDW) state
\begin{equation} \label{def_cdw_state}
|\Psi_0\rangle = |\rm{CDW} \rangle\,,
\end{equation}
in which each occupied single-particle state $|j\rangle$ has only unoccupied neighbors, and hence the particle filling is $N/D=1/2$.
{This type of CDW state is considered as the ``chess'' state in Fig.~\ref{sketch}.
Unless explicitly mentioned otherwise, all further numerical results refer to the "chess" CDW state.
}%
In Sec.~\ref{sec:observables} we then extend our analysis to the other types of CDW states from Fig.~\ref{sketch}.
The class of initial states that we consider can be generally written as
\begin{equation}
    |\Psi_0\rangle = \prod_{j_l  \in \Psi_0} \hat c_{j_l}^\dagger |\emptyset\rangle \;,
\end{equation}
where $j_l \in \Psi_0$ denotes the occupied sites.

Since the observable $\hat{n}_{j}$ shares the same eigenbasis with the initial Hamiltonian $\hat H_0$, one can apply Eqs.~(\ref{def_oA_short}) and~(\ref{def_oA_final}) to describe their dynamics.
Specifically, we distinguish between the dynamics of occupations of initially occupied sites, 
\begin{equation} \label{def_ni_1}
n^{\rm occ}_j(t) \equiv n_{j\in \Psi_0}(t) = P_{jj}(t) + \sum_{\substack{j_l \in \Psi_0}}^{j_l\neq j}  P_{jj_l}(t)\;,
\end{equation}
for which $n^{\rm occ}_j(t=0)=1$,
and of occupations of initially unoccupied (empty) sites,
\begin{align} \label{def_ni_2}
n^{\rm unocc}_j(t) \equiv n_{j\notin \Psi_0}(t) = \sum_{\substack{j_l \in \Psi_0}}  P_{jj_l}(t)\;,
\end{align}
for which $n^{\rm unocc}_j(t=0)=0$.
The interpretation of Eqs.~(\ref{def_ni_1}) and~(\ref{def_ni_2}) is that the sums represent the flows of particles from sites $j_l$ to the site $j$.

To simplify the analysis, we only study the averages over the occupied and unoccupied sites, respectively, denoted as $n_{\rm occ}(t)$ and $n_{\rm unocc}(t)$.
They are defined as
\begin{align}
n_{\rm occ}(t) &=\big\langle\big\langle\, n^{\rm occ}_j(t) \,\big\rangle_{j\in\Psi_0}\, \big\rangle_H \;,
\nonumber\\
n_{\rm unocc}(t) &=\big\langle\big\langle\, n^{\rm unocc}_j(t) \,\big\rangle_{j\notin\Psi_0}\, \big\rangle_H \;, \label{eq:def_nt}
\end{align}
where $\braket{\dots}_{j\in\Psi_0}$ ($\braket{\dots}_{j\notin\Psi_0}$) denotes the average over all initially occupied (unoccupied) sites, while $\braket{\dots}_H$ denotes the average over different Hamiltonian realizations.
They satisfy the sum rule
\begin{equation}
    n_{\rm occ}(t) + n_{\rm unocc}(t) = 1 \;.
\end{equation}
Moreover, they also allow for a simple definition of the imbalance,
\begin{align} \label{eq:def_imbalance}
    I(t) = n_{\rm occ}(t) - n_{\rm unocc}(t) = 2n_{\rm occ}(t)-1\;,
\end{align}
which equals $I(t=0)=1$ and approaches $I(t\to\infty) \to 0$ if the state at long times is completely delocalized.
In what follows, we mostly focus on the dynamics of the imbalance.

\begin{figure}[!t]
\includegraphics[width=0.98\columnwidth]{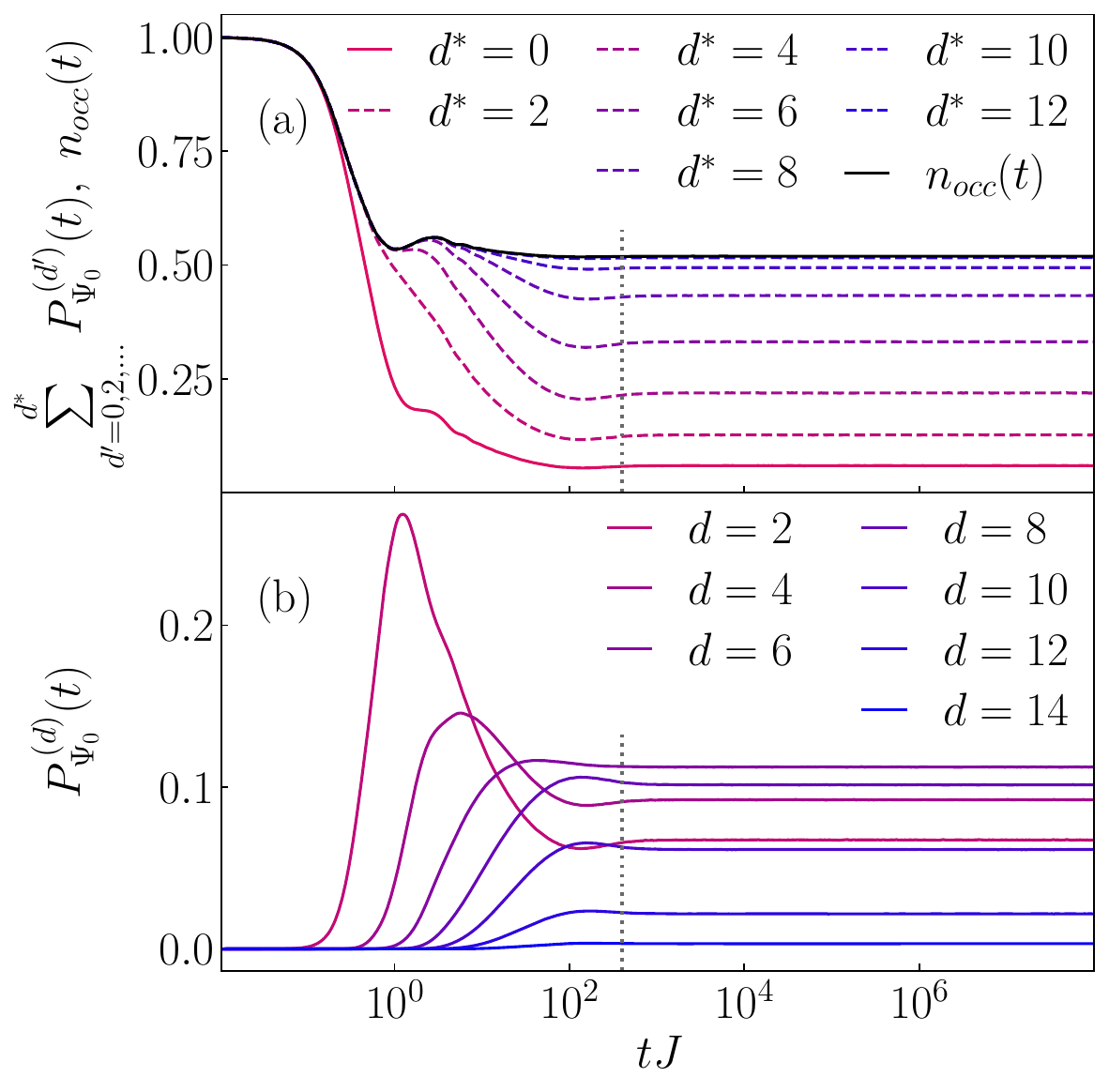}
\vspace{-0.2cm}
\caption{
Contributions of single-particle transition probabilities $P_{\Psi_0}^{(d)}(t)$ from Eq.~(\ref{def_Ppsi_d_1}) to the dynamics of occupied sites $n_{\rm occ}(t)$ for the initial CDW state for the 3D Anderson model with $L=10$ and at the critical point $W_c/J=16.5$.
(a) Sums of the contributions at distances up to a maximal distance $d^*$.
(b) Separate contributions at a given $d$.
The vertical grey lines in both panels indicate the typical Heisenberg time $t_\mathrm{H}^{\mathrm{typ}}$ from Eq.~(\ref{def_tHeis}). 
}
\label{fig1}
\end{figure}

\begin{figure}[!t]
\includegraphics[width=0.98\columnwidth]{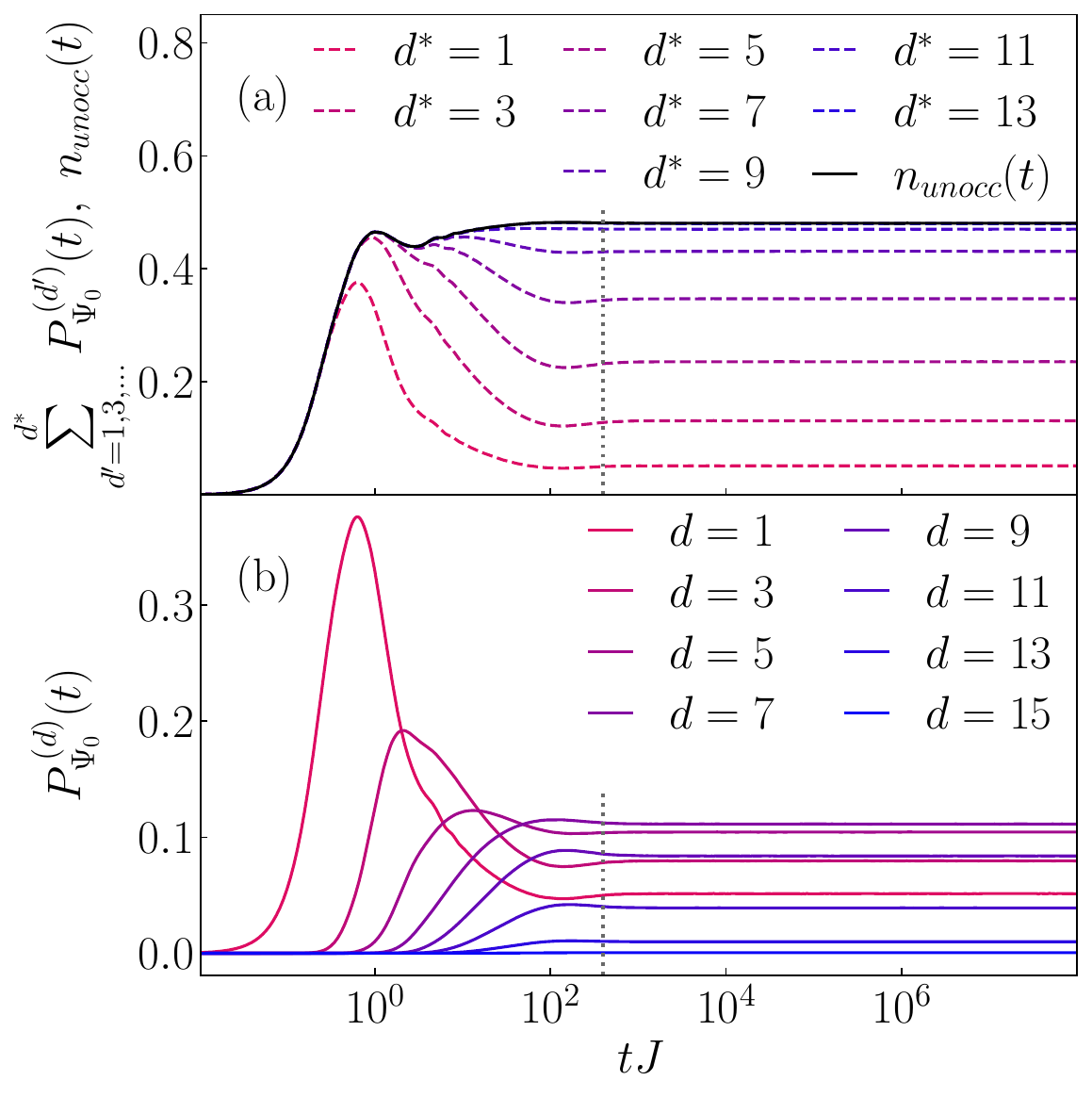}
\vspace{-0.2cm}
\caption{
Contributions of single-particle transition probabilities $P_{\Psi_0}^{(d)}(t)$ from Eq.~(\ref{def_Ppsi_d_2}) to the dynamics of unoccupied sites $n_{\rm unocc}(t)$ for the initial CDW state for the 3D Anderson model with $L=10$ and at the critical point $W_c/J=16.5$.
(a) Sums of the contributions at distances up to a maximum distance $d^*$.
(b) Separate contributions at a given $d$.
The vertical grey lines in both panels indicate the typical Heisenberg time $t_\mathrm{H}^{\mathrm{typ}}$ from Eq.~(\ref{def_tHeis}).
}
\label{fig2}
\end{figure}

Before proceeding, we discuss another perspective on the dynamics of site occupations.
For the case of the initial ``chess'' CDW state, the contributions to $n_{\rm occ}(t)$ come from the initially occupied single-particle states $j_l$ for which the distance $d$ to any other initially occupied site $j$ is an even number. Note that the distance $d = ||\mathbf{r}_j - \mathbf{r}_{j_l}||_1$ is given by the $\mathcal{\ell}_1$-norm, i.e., the minimum number of hops between the sites.
We express this property as
\begin{align} \label{eq:def_densities_1}
n_{\rm occ}(t) = \sum_{d' = 0,2,\dots} P_{\Psi_0}^{\,(d')}(t) \;,
\end{align}
where the cumulative transition probabilities to the initially occupied sites at {\it even} distance $d$ are defined as
\begin{align} \label{def_Ppsi_d_1}
P_{\Psi_0}^{\,(d)}(t)= \Bigg<\Bigg<\sum_{\substack{j_l\in \Psi_0 \\ ||\mathbf{r}_j - \mathbf{r}_{j_l}||_1 = d}}  P_{jj_l}(t) \Bigg>_{j\in\Psi_0}\,\Bigg>_{H}\;.
\end{align}

Analogously, the contributions to $n_{\rm unocc}(t)$ come from the initially occupied single-particle states $j_l$ at odd distances $d$,
\begin{align} \label{eq:def_densities_2}
n_{\rm unocc}(t) = \sum_{d' = 1,3,\dots} P_{\Psi_0}^{\,(d')}(t)\;,
\end{align}
where the cumulative transition probabilities to the initially unoccupied sites at {\it odd} distance $d$ are defined as
\begin{align} \label{def_Ppsi_d_2}
P_{\Psi_0}^{\,(d)}(t)= \Bigg<\Bigg<\sum_{\substack{j_l\in \Psi_0 \\ ||\mathbf{r}_j - \mathbf{r}_{j_l}||_1 = d}}  P_{jj_l}(t) \Bigg>_{j\notin\Psi_0}\,\Bigg>_{H}\;.
\end{align}
In Fig.$~$\ref{fig1}(a), we show the dynamics of site occupations $n_{\rm occ}(t)$ and the sum of the contributions from transition probabilities $P_{\Psi_0}^{(d')}(t)$ at even distances $d'$ up to a maximum value $d'=d^*$.
We study the 3D Anderson model at the critical point $W_c/J=16.5$.
We see that, as expected, by increasing $d^*$ one approaches $n_{\rm occ}(t)$. 
Analogous results are shown in Fig.$~$\ref{fig2}(a) for the site occupations $n_{\rm unocc}(t)$ and the sum of the contributions from transition probabilities $P_{\Psi_0}^{(d')}(t)$ at odd distances $d'$ up to a maximum value $d'=d^*$.

In Figs.~\ref{fig1}(b) and~\ref{fig2}(b), we show the independent contributions from $P_{\Psi_0}^{(d)}(t)$ at even and odd $d$, respectively.
As argued below, $P_{\Psi_0}^{(d)}(t)$ at $d=0$ is essentially identical to the survival probability, for which the scale invariance at the criticality was established in Ref.$~$\cite{hopjan2023}.
At $d>0$ the transition probabilities start from zero at $t=0$ and exhibit a peak at $t>0$, which is followed by the decay to their infinite-time value. 
In the next section we explore whether the notion of scale invariance carries over to the dynamics of transition probabilities.


\section{Scale invariance of transition probabilities}\label{sec:transition_prob}

\begin{figure}[!t]
\includegraphics[width=\columnwidth]{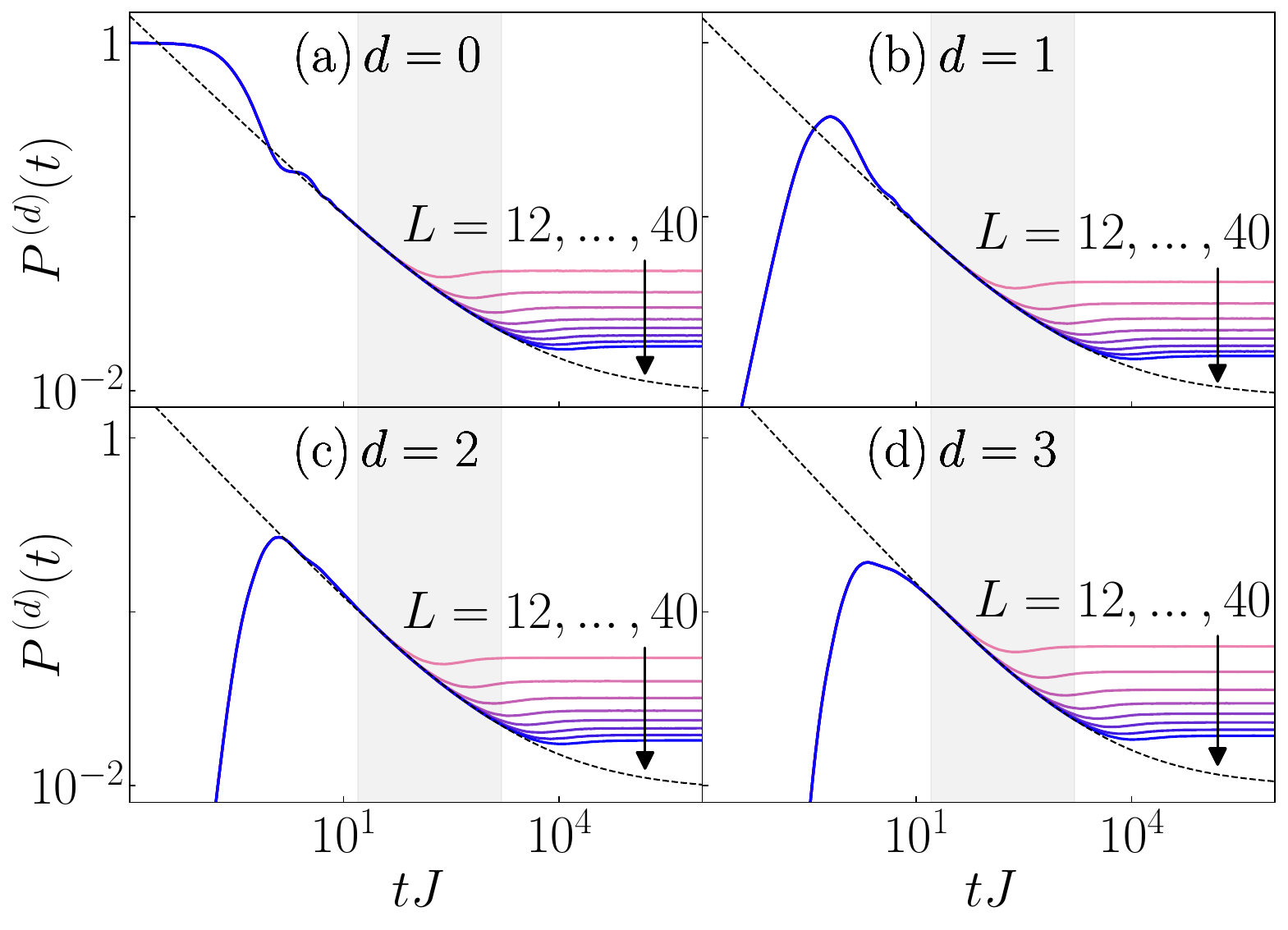}
\vspace{-0.2cm}
\caption{
Time evolution of transition probabilities $P^{\,(d)}(t)$ from Eq.~(\ref{eq:def_Pd}) for $d=0,1,2,3$ and system sizes $L = 12,16,20,24,28,32,36,40$, for the 3D Anderson model at the critical point $W_c/J=16.5$.
The dotted lines are three-parameter fits (within the shaded region at $L=40$) to a function $a (tJ)^{-\beta}+P^{\,(d)}_\infty$, where $P^{\,(d)}_\infty>0$.
}
\label{fig3}
\end{figure}

{The case study in Sec.~\ref{sec:case_study} motivates us to consider more closely the behavior of the single-particle transition probabilities. 
They govern the dynamics of site occupations in many-body states, such as those given by Eq.~(\ref{eq:def_densities_1}).
}%

{
We first note a technical detail that the transition probabilities at a given $d$, in the case of the initial CDW state, were only averaged over a fraction of all lattice sites. More precisely, these averages were taken over $N=D/2$ sites for even $d$ and $D-N=D/2$ sites for odd $d$. Contrary, in this section, we define the transition probabilities with the average taken over all lattice sites. This type of averaging is motivated by two reasons.
We want, first, to study the averages that are independent of the particular form of the initial many-body state, and, additionally, to connect to the previous study of survival probabilities where the averaging over all sites was used, (see Ref.$~$\cite{hopjan2023}). 
Nevertheless, below we discuss that the differences between these types of averages are very small.}

\begin{figure}[!t]
\includegraphics[width=\columnwidth]{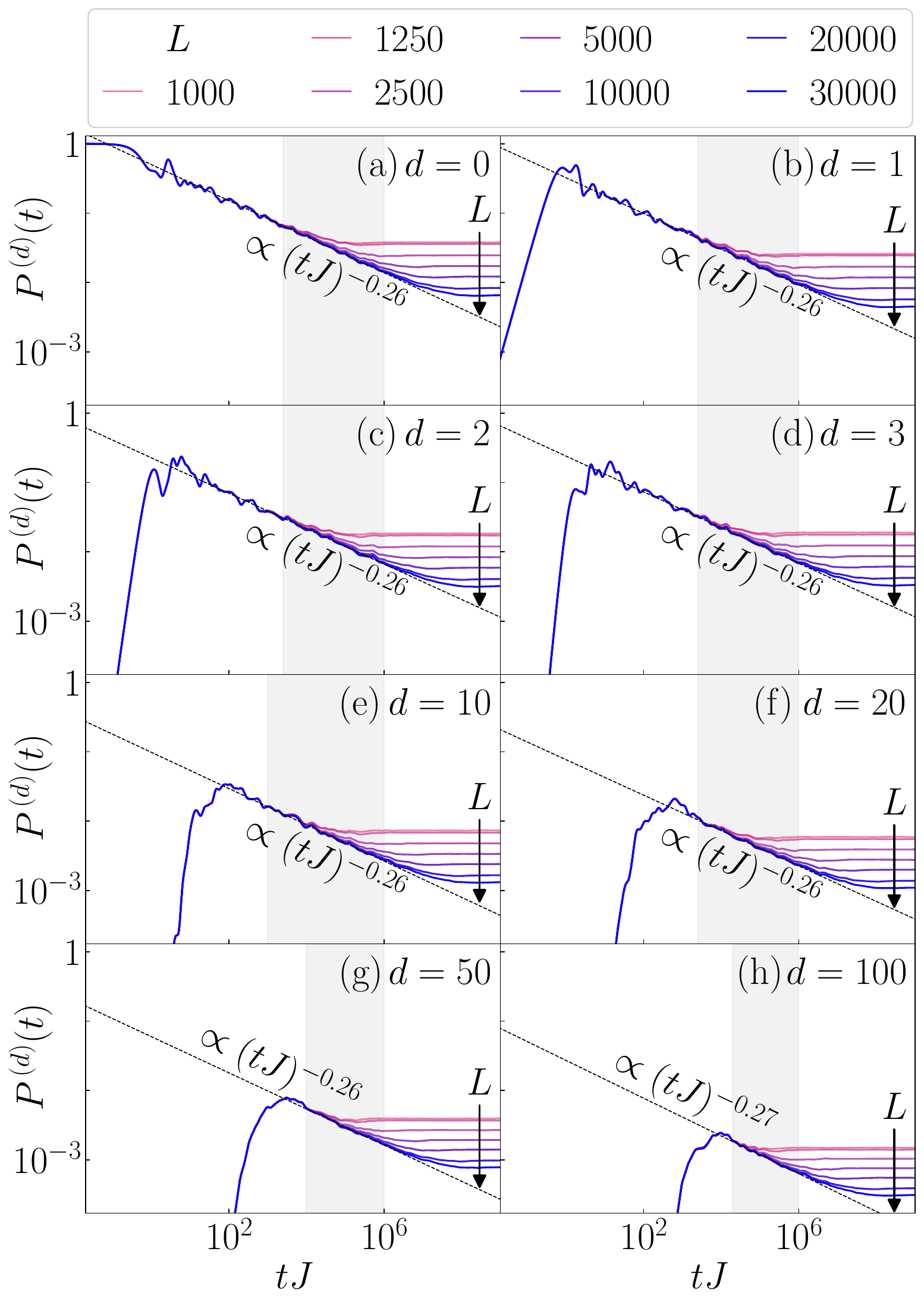}
\vspace{-0.2cm}
\caption{
Time evolution of transition probabilities $P^{\,(d)}(t)$ from Eq.~(\ref{eq:def_Pd}) for different values of $d$ and system sizes $L$, as indicated in the legend, for the 1D Aubry–André model at the critical point $\lambda_c/J=2$.
The dotted lines are two-parameter fits (within the shaded region at $L=30000$) to a function $a (tJ)^{-\beta}$.
}
\label{fig4}
\end{figure}

{The central goal of this section is to show that the transition probabilities exhibit scale-invariant critical dynamics analogous to the behavior of the survival probabilities from Ref.~\cite{hopjan2023}.
We define the generalized averages of the transition probabilities as}
\begin{align}
    \label{eq:def_Pd}
    P^{\,(d)}(t) = \braket{\,\braket{\,P_j^{\,(d)}(t)\,}_j}_H \;,
\end{align}
where $P_j^{\,(d)}(t)=\sum_{k,||\mathbf{r}_k - \mathbf{r}_{j}||_1 = d}P_{kj}(t)$ measures the cumulative transition probability from the initially localized single-particle state at site $j$ to all sites at distance $d$, and $\braket{\dots}_j$ denotes the average over all sites $j$. 
Alternatively, the transition probability $P_j^{\,(d)}(t)$ can also be interpreted as the inflow of particles from a distance $d$ to the site $j$, assuming that in the initial state all sites at  distance $d$ are occupied.
The limiting case $d=0$ then corresponds to the averaged survival probability studied in Ref.$~$\cite{hopjan2023}.
{We note, however, that the differences between $P^{(d)}(t)$ from Eq.~(\ref{eq:def_Pd}) and $P_{\Psi_0}^{(d)}(t)$ from Eqs.~(\ref{def_Ppsi_d_1}) and~(\ref{def_Ppsi_d_2}) are negligible (see Appendix$~$\ref{sec:averaging}), at least for the system sizes under investigation. Therefore, we only focus on $P^{(d)}(t)$ in the remainder of the paper.}

In Fig.$~$\ref{fig3}, we plot the transition probabilities $P^{\,(d)}(t)$ for the 3D Anderson model at distances $d=0,1,2,3$, and in Fig.$~$\ref{fig4}, we plot the corresponding results for the 1D Aubry–André model at distances $d=0,1,2,3,10,20,50,100$. 
At $d=0$, the results coincide with the survival probability and hence they exhibit a power-law decay to the infinite-time value that decreases with $L$~\cite{hopjan2023}.
At $d>0$, the transition probabilities start from zero and then exhibit a peak.
At later times, they exhibit power-law decays with similar properties as those at $d=0$.
At $d \gg 1$, however, the peak in transition probabilities emerges at times $tJ \gg 1$, so that the time range of the power-law decay shrinks accordingly [see the shrinking of the shaded regions in Figs.~\ref{fig4}(f)-\ref{fig4}(h)].

\begin{figure}[!t]
\includegraphics[width=\columnwidth]{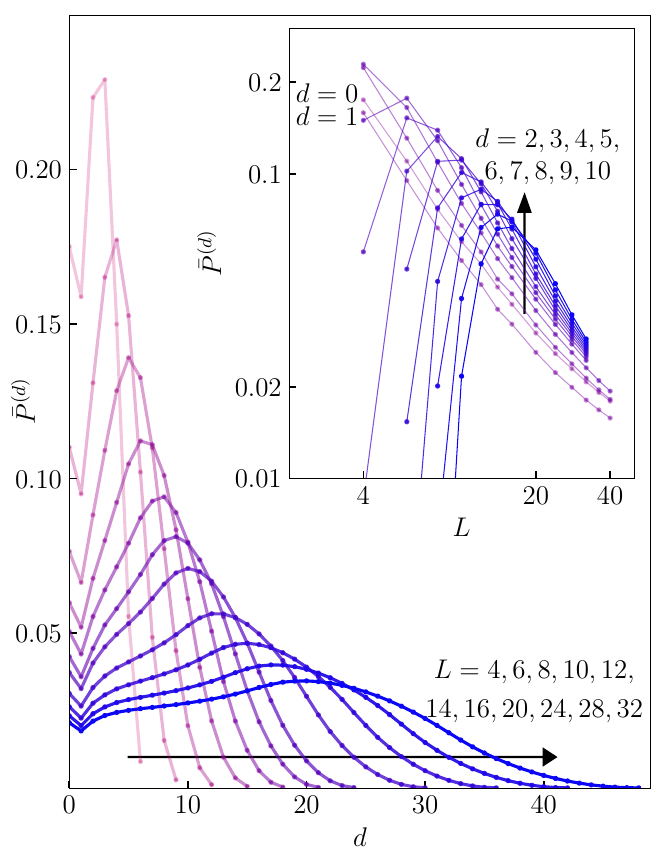}
\vspace{-0.4cm}
\caption{
Properties of the infinite-time values $\overline{P}^{(d)}$ from Eq.~(\ref{def_Pbar}) in the 3D Anderson model at the critical point $W_c/J=16.5$.
Main panel: $\overline{P}^{(d)}$ as a function of the distance $d$ for several system sizes $L$.
Inset: $\overline{P}^{(d)}$ as a function of $L$ for several values of $d$.
Results are averaged over 50 realizations of the Hamiltonian.
}
\label{fig6}
\end{figure}

\begin{figure}[!t]
\includegraphics[width=\columnwidth]{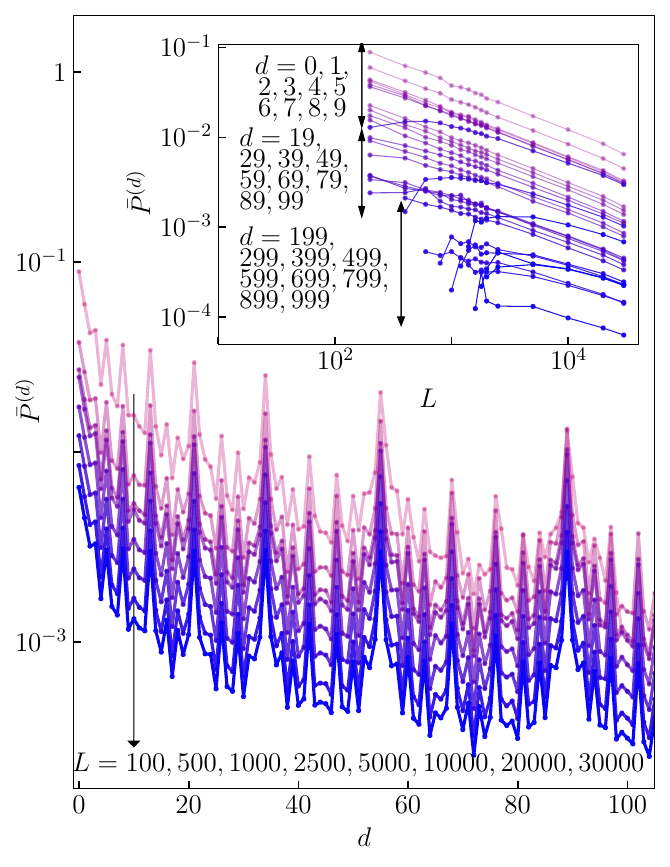}
\vspace{-0.4cm}
\caption{
Properties of the infinite-time values $\overline{P}^{(d)}$ from Eq.~(\ref{def_Pbar}) in the 1D Aubry–André model at the critical point $\lambda_c/J=2$.
Main panel: $\overline{P}^{(d)}$ as a function of the distance $d$ for several system sizes $L$.
Inset: $\overline{P}^{(d)}$ as a function of $L$ for several values of $d$.
} 
\label{fig7}
\end{figure}

\subsection{Long-time behavior of transition probabilities}

Next, we study the infinite-time behavior of $P^{(d)}(t)$.
We define the infinite-time value at a fixed system size $\overline{P}^{(d)}$ and the infinite-time value in the thermodynamic limit $P_\infty^{(d)}$, respectively, as
\begin{equation} \label{def_Pbar}
    \overline{P}^{(d)} = \lim_{t\to\infty} P^{(d)}(t)\;,\;\;\;
    P_\infty^{(d)} = \lim_{D\to\infty} \overline{P}^{(d)} \;.
\end{equation}
Note that $\overline{P}^{(0)}$ represents the inverse participation ratio (IPR) in the single-particle eigenbasis of the final Hamiltonian $\hat{H}$,
\begin{equation} \label{def_ipr}
\overline{P}^{(0)} = \langle\langle \sum_q |c_{qj}|^4\rangle_j\rangle_H\;,
\end{equation}
where $c_{qj} = \langle q|j\rangle$ is the overlap of the single-particle eigenstate $|j\rangle$ of the initial Hamiltonian $\hat{H}_{0}$ (or the one-body observable $\hat{o}$) with the single-particle eigenstate $|q\rangle$ of the Hamiltonian $\hat H$ after the quench.

Figures~\ref{fig6} and~\ref{fig7} show $\overline{P}^{(d)}$ for the 3D Anderson model and the 1D Aubry–André model, respectively, as a function of $d$ at different $L$ (main panels) and as a function of $L$ at different $d$ (insets).
In the 3D Anderson model, $\overline{P}^{\,(d)}$ is a smooth function of $d$ with a maximum at $d\approx d_{\rm max}/2$ (see the main panel of Fig.~\ref{fig6}; $d_max$ is the maximal distance between pairs of sites in a given lattice).
The shape of the profile is governed by the number of available sites at distance $d$.
At $d=0$, $\overline{P}^{\,(d)}$ exhibits a power-law decay in $L$ to a nonzero value $P_\infty^{(d)}>0$, which is a consequence of the mobility edge at the critical point of the 3D Anderson model~\cite{Schubert_05}.
This property manifests itself on a log-log scale in the inset of Fig.~\ref{fig6} (for $d=0$) as an upward bending of the curve at large $L$.
Results for larger distances $d>0$ also suggest the emergence of a nonzero asymptotic value $P_\infty^{(d)}>0$ and hence we model the approach to the thermodynamic limit by the ansatz
\begin{align}
    \label{eq:def_Pd_infty}
   \overline{P}^{\,(d)} = c_d D^{-\gamma_d} + P_\infty^{\,(d)}\;.
\end{align}
At $d=0$, $\gamma_0 \equiv \gamma$ is interpreted as the fractal dimension, since it corresponds to the decay of the IPR from Eq.~(\ref{def_ipr}) upon increasing the Hilbert-space dimension $D$.
At sufficiently small $d$, we get $\gamma_d\approx\gamma \approx 0.4$ (see the results for $d=0,1,2,3$ in the insets of Fig.~\ref{fig8}).
Note, however, that the ansatz in Eq.~(\ref{eq:def_Pd_infty}) should only be applicable in the asymptotic regime, i.e., at sufficiently large $L>L^*$.
In general, $L^*$ increases with the distance $d$. In a given finite system, the distances $d \gtrsim d_{\rm max}/2$ are not yet in the asymptotic regime.

In the 1D Aubry–André model, $\overline{P}^{\,(d)}$ is not a smooth function of $d$ (see the main panel of Fig.$~$\ref{fig7}). It exhibits resonances at certain distances $d$.
Simultaneously, $\overline{P}^{\,(d)}$ appears to decay as a power law towards zero with $L$ for all $d$ (see the inset of Fig.$~$\ref{fig7}). Therefore, the thermodynamic-limit value vanishes, $P^{\,(d)}_\infty=0$ in Eq.~(\ref{eq:def_Pd_infty}), which is expected at $d=0$~\cite{hopjan2023}. We note, however,  that at certain distances $d$, the asymptotic regime is realized for values of $L$ that are several orders of magnitude larger than $d$. 
Nonetheless, we conjecture that the ansatz in Eq.~(\ref{eq:def_Pd_infty}) with $P^{\,(d)}_\infty=0$ is applicable for all $d$ in the thermodynamic limit.
Specific examples of the fits to Eq.~(\ref{eq:def_Pd_infty}) are shown in the insets of Fig.~\ref{fig9} for $d=0,1,2,3,10,20,50,100$, and we obtain $\gamma_d \approx 0.5$ in all cases.

\subsection{Emergence of scale invariance}

Motivated by Refs.~\cite{hopjan2023, hopjan2023scaleinvariant}, we now introduce the scaled transition probabilities $p^{(d)}(\tau)$, defined as
\begin{align}
    \label{eq:def_resc_prop}
    p^{(d)}(\tau) = \frac{P^{\,(d)}(\tau) - P_\infty^{\,(d)}}{\overline{P}^{\,(d)} - P_\infty^{\,(d)}} \;,
\end{align}
where the scaled time $\tau$ is measured in units of the typical Heisenberg time $t_H^\mathrm{typ}$,
\begin{align} \label{def_tHeis}
    \tau=t/t_H^\mathrm{typ}\;,\;\;\; t_H^\mathrm{typ} = 2\pi\,\text{e}^{-\braket{\braket{\ln(\varepsilon_{q+1} - \varepsilon_q)}_q}_H} \;,
\end{align}
in which $\braket{\dots}_q$ denotes the average over all neighboring energy levels $\varepsilon_q$ and $\braket{\dots}_H$ is the average over Hamiltonian realizations.

\begin{figure}[!t]
\includegraphics[width=\columnwidth]{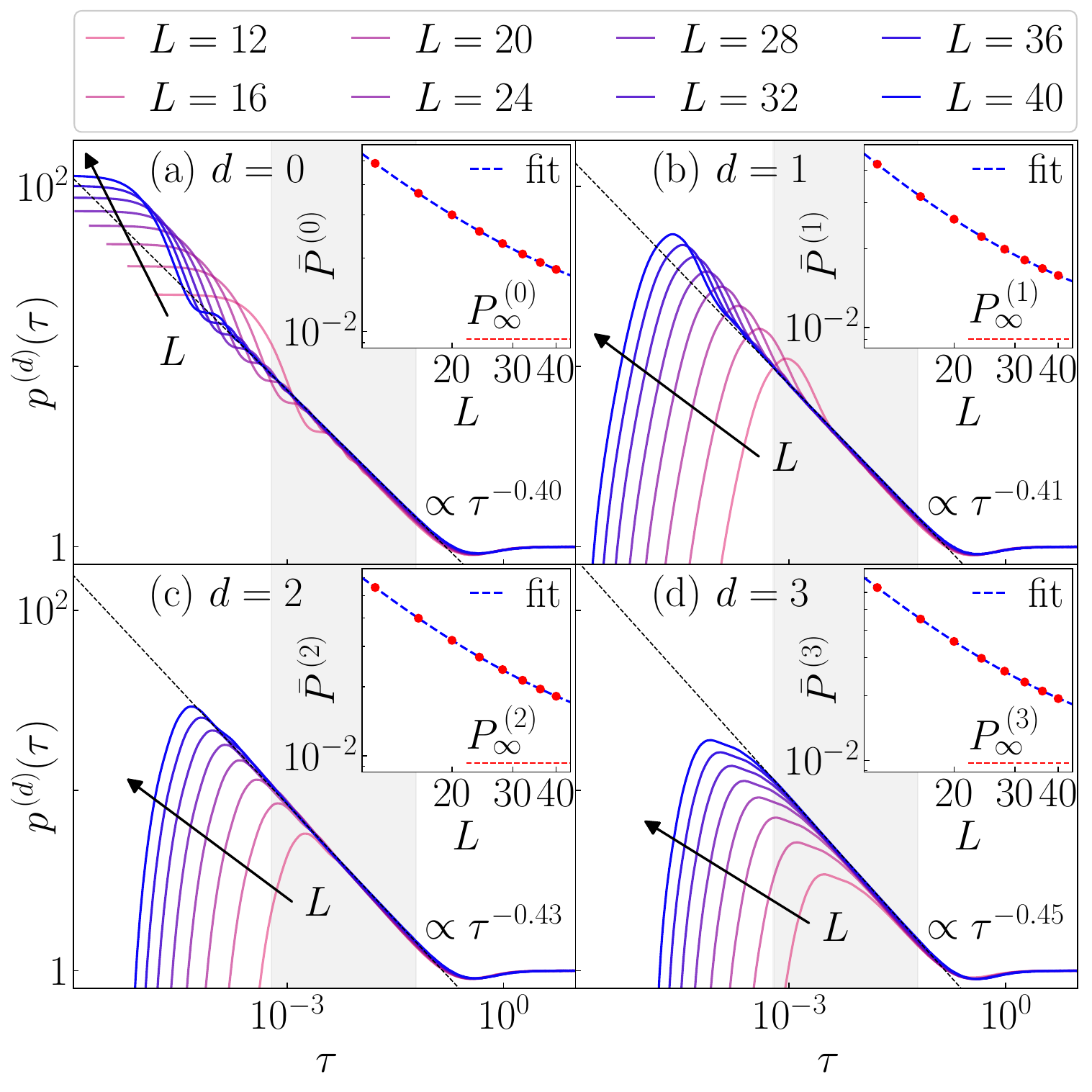}
\vspace{-0.2cm}
\caption{
Main panels: Transition probabilities $p^{(d)}(\tau)$ from Eq.~(\ref{eq:def_resc_prop}) for the 3D Anderson model at the critical point $W_c/J=16.5$ as a function of scaled time $\tau$ and different system sizes $L$ [the same data as in Fig.$~$\ref{fig3}].
Dashed lines are two-parameter fits to Eq.~(\ref{eq:scaling_ansatz}), where the shaded area corresponds to the parameter range used in the fit. 
Insets: Infinite-time values $\overline{P}^{(d)}$ versus $L$ [circles] and the three-parameter fits to Eq.~(\ref{eq:def_Pd_infty}) [dashed lines].
The horizontal dotted lines are the infinite-time values in the thermodynamic limit $P_\infty^{(d)}$.
Results are shown for (a) $d=0$, (b) $d=1$, (c) $d=2$ and (d) $d=3$.
}
\label{fig8}
\end{figure}

\begin{figure}[!t]
\includegraphics[width=\columnwidth]{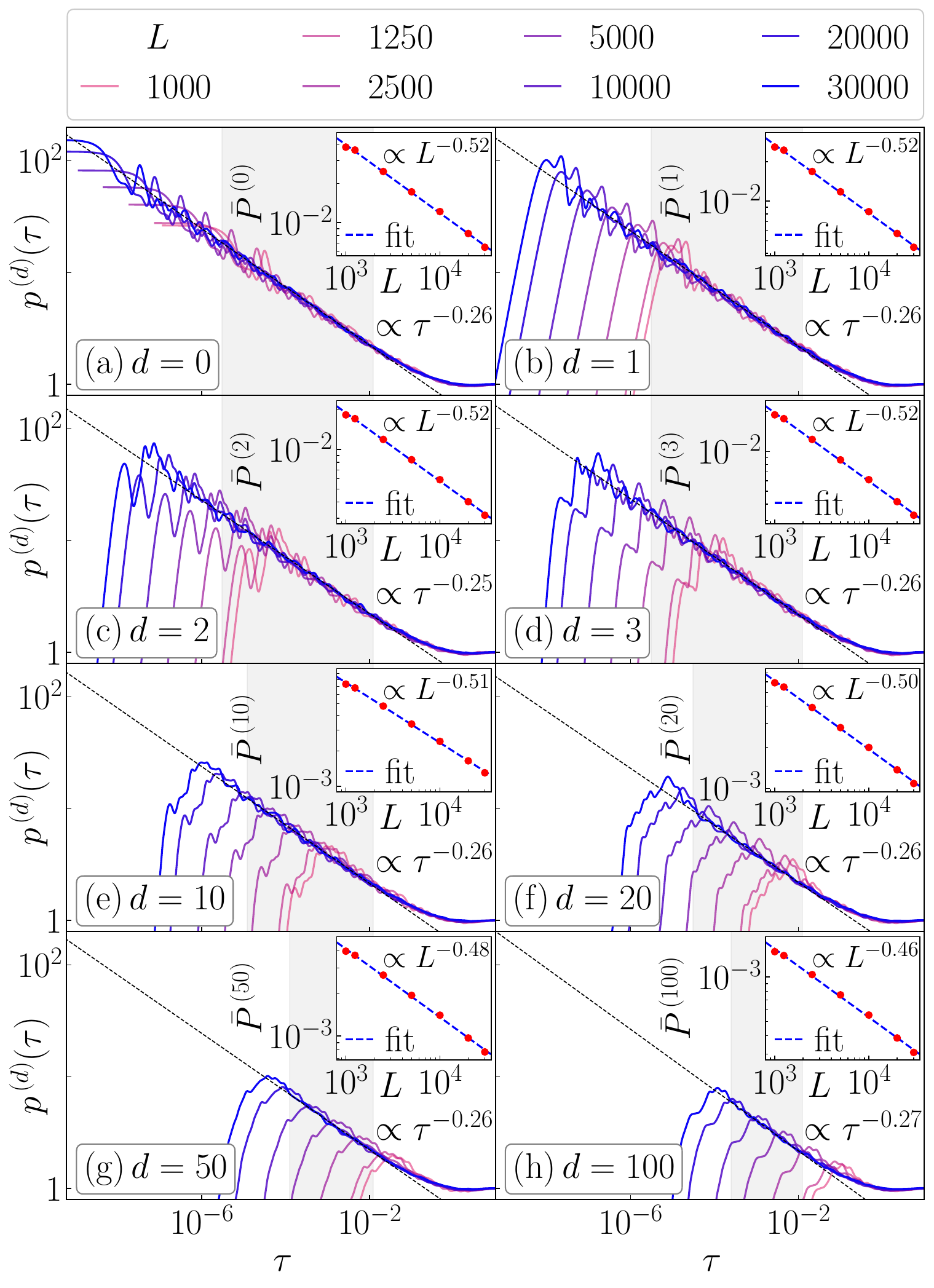}
\vspace{-0.2cm}
\caption{
Main panels: Transition probabilities $p^{(d)}(\tau)$ from Eq.~(\ref{eq:def_resc_prop}) in the 1D Aubry–André model at the critical point $\lambda_c/J=2$ as a function of scaled time $\tau$ and different system sizes $L$ [the same data as in Fig.$~$\ref{fig4}].
The dashed lines are two-parameter fits to Eq.~(\ref{eq:scaling_ansatz}), where the shaded area corresponds to the fitted range. 
Insets: Infinite-time values $\overline{P}^{(d)}$ versus $L$ [circles] and the two-parameter fits to Eq.~(\ref{eq:def_Pd_infty}), setting $P_\infty^{(d)}=0$ [dashed lines].
Results range from (a) $d=0$ to (h) $d=100$, as indicated in the legends. To smoothen the results, we used a Gaussian filter function with a standard deviation of 5 data points.
}
\label{fig9}
\end{figure}

The results for $p^{(d)}(\tau)$ are shown in Fig.$~$\ref{fig8} at the critical point of the 3D Anderson model, and in  Fig.$~$\ref{fig9} at the critical point of the 1D Aubry–André model. 
We observe the emergence of a scale-invariant decay in all cases under consideration.
This is the main result of this section and can be considered as an extension of the concept of the scale-invariant survival probability ($d=0$)~\cite{hopjan2023, hopjan2023scaleinvariant} to the transition probabilities ($d>0)$.

We highlight two properties of the scale-invariant transition probabilities.
First, the onset time of the scale invariance is expected to be much shorter than the Heisenberg time.
As argued in Ref.~\cite{hopjan2023scaleinvariant} for $d=0$, while the dynamics close to the Heisenberg time can be referred to as the {\it late-time} dynamics, the scale invariance emerges already in the {\it mid-time} dynamics.
Note that similar onset times of the scale invariance are observed in Figs.~\ref{fig8} and~\ref{fig9} at nonzero values of $d$ that are not too large.
For large $d$, however, the onset times of scale invariance become long, so that the time interval of scale invariance in finite systems shrinks [see the shaded regions in Figs.~\ref{fig9}(f)--\ref{fig9}(h)].
In these cases, the distinction between the mid-time and late-time dynamics becomes less clear.
Nevertheless, we still expect $p^{(d)}(\tau)$ at all $d$ to become scale invariant at sufficiently large time and system size.

The second property is that the scale-invariant dynamics exhibits a power-law decay, which we describe by
\begin{align}
    \label{eq:scaling_ansatz}
    p^{\,(d)}(\tau) = a_d\,\tau^{-\beta_d}\;.
\end{align}
The power-law decay is a consequence of the choice of the initial state, which is fractal in the eigenbasis at the critical points of both Hamiltonians under investigation~\cite{Ketzmerick_92,Huckestein94,Brandes96,Ketzmerick_97,Ohtsuki97,hopjan2023,hopjan2023scaleinvariant}.
The decay exponents $\beta_d$ in the 3D Anderson model in Fig.$~$\ref{fig8} lie in the interval $\beta_d \in [0.40,0.45]$, while for the 1D Aubry–André model (see Fig.$~$\ref{fig9}), they are essentially the same, $\beta_d\approx 0.26$. 
It is beyond the scope of this work to provide a rigorous answer to the question whether $\beta_d$ should be identical for all $d$ within a given model.

\section{Critical behavior of observables}  \label{sec:observables}

We now turn our attention back to the quench dynamics of the initial CDW states of interest and we combine the results from Secs.~\ref{sec:case_study} and~\ref{sec:transition_prob} to study the critical dynamics of observables.

We first focus on the dynamics of site occupations of initially occupied sites $n_{\rm occ}(t)$ [see Eq.$~$\eqref{eq:def_densities_1}], for the initial CDW state $|\Psi_0\rangle$ from Eq.~(\ref{def_cdw_state}).
We simplify the notation $n_{\rm occ}(t) \to n(t)$ further on.
Expressing the transition probabilities $P^{(\delta)}$ via Eq.$~$\eqref{eq:def_resc_prop}, we rewrite Eq.$~$\eqref{eq:def_densities_1} as
\begin{align}
    \label{eq:n_t_p_inserted}
    n(t) = \hspace*{-0.3cm} \sum_{\delta=0,2,4,...}\hspace*{-0.1cm}\left[P_\infty^{\,(\delta)} + p^{\,(\delta)}(t)\left(\,\overline{P}^{\,(\delta)} - P_\infty^{\,(\delta)}\right)\right].
\end{align}
The first quantity to be identified in Eq.~(\ref{eq:n_t_p_inserted}) is the asymptotic value of the infinite-time site occupations,
\begin{equation} \label{def_n_infty}
n_\infty=\lim\limits_{D\xrightarrow{}\infty}\,\lim\limits_{t\xrightarrow{}\infty}\,n(t) = \sum\limits_{\delta=0,2,4,...}\hspace*{-0.3cm} P_\infty^{\,(\delta)} \;,
\end{equation}
which simplifies Eq.~\eqref{eq:n_t_p_inserted} to
\begin{align}
    n(t) - n_\infty = \sum_{\delta=0,2,4,...}\hspace*{-0.1cm}p^{\,(\delta)}(t)\left(\,\overline{P}^{\,(\delta)} - P_\infty^{\,(\delta)}\right)\;.
\end{align}
In the next step, we substitute the time $t$ with $\tau$ and we replace $p^{(\delta)}(t)$ with the power-law ansatz from Eq.~\eqref{eq:scaling_ansatz}, yielding
\begin{align} \label{eq:n_tau_2}
    n(\tau) - n_\infty = \sum_{\delta=0,2,4,...}\hspace*{-0.1cm} a_{\delta}\,\tau^{-\beta_{\delta}}\left(\,\overline{P}^{\,(\delta)} - P_\infty^{\,(\delta)}\right).
\end{align}
At this stage, the scale-invariant dynamics is not yet guaranteed, since $\overline{P}^{(\delta)}$ in Eq.~(\ref{eq:n_tau_2}) depends on the system size.
Below, we discuss scenarios in which the scale-invariant dynamics of observables may emerge.

\subsection{Scale-invariant dynamics of observables}

A simplification of Eq.~(\ref{eq:n_tau_2}) generally requires the knowledge of all coefficients $a_{\delta}$, $\beta_{\delta}$.
We first consider an approximation, in which the coefficients do not depend on $\delta$, i.e., we replace $a_{\delta} \to a$ and $\beta_{\delta} \to \beta$.
In this case, one obtains
\begin{align} \label{eq:n_approximate}
    n(\tau) - n_\infty = a\,\tau^{-\beta}\sum_{\delta=0,2,4,...}\hspace*{-0.1cm}\left(\,\overline{P}^{\,(\delta)} - P_\infty^{\,(\delta)}\right), 
\end{align}
where we can identify the infinite-time value of site occupations,
\begin{equation} \label{def_n_average}
\overline{n} = \lim\limits_{t \xrightarrow{}\infty}n(t) =  \sum_{\delta=0,2,4,...}\hspace*{-0.1cm}\overline{P}^{\,(\delta)}\;,
\end{equation}
which equals the GGE prediction, $\overline{n} = n_{\rm GGE}$~\cite{rigol_dunjko_07, vidmar16, Ziraldo_2012,He_2013, Lydzba23}.
This motivates us to introduce the scaled site occupations $\tilde{n}$, 
\begin{align}\label{eq:n_scaled}
   \tilde{n}(\tau)= \frac{n(\tau) - n_\infty}{\overline{n}  - n_\infty}\;,
\end{align}
which we expect to exhibit scale-invariant dynamics when measured as a function of $\tau$.
In particular, using the approximation from Eq.~(\ref{eq:n_approximate}), one obtains the prediction
\begin{align}\label{eq:n_analytic}
   \tilde{n}(\tau) = a\,\tau^{-\beta}\;.
\end{align}
Analogously, one can also introduce the scaled imbalance,
\begin{align}\label{imbalance_rescaled}
    \tilde{I}(\tau) = \frac{I(\tau) - I_\infty}{\overline{I} - I_\infty} \;,
\end{align}
where the asymptotic value $I_\infty$ is the analog of $n_\infty$ from Eq.~(\ref{def_n_infty}), and the infinite-time value $\overline{I}$ is the analog of $\overline{n}$ from Eq.~(\ref{def_n_average}).
Using the relationship between the site occupations and imbalance from Eq.~(\ref{eq:def_imbalance}), and the approximation from Eq.~(\ref{eq:n_approximate}), one obtains the prediction
\begin{align}\label{imbalance_approximate}
    \tilde{I}(\tau) = a\,\tau^{-\beta} \;,
\end{align}
which is identical to the one for site occupations in Eq.~(\ref{eq:n_analytic}).

We highlight that the approximation from Eq.~(\ref{eq:n_approximate}), in which one replaces the time dependence of all transition probabilities with the same functional form, may be too strong and at this point, we cannot provide a rigorous argument for its justification.
Here we only provide numerical justification based on the results from Figs.~\ref{fig8} and~\ref{fig9}, in which we observe very similar dynamical properties of the transition probabilities in the 3D Anderson and the 1D Aubry–André model, respectively.

\begin{figure}[!t]
\includegraphics[width=0.99\columnwidth]{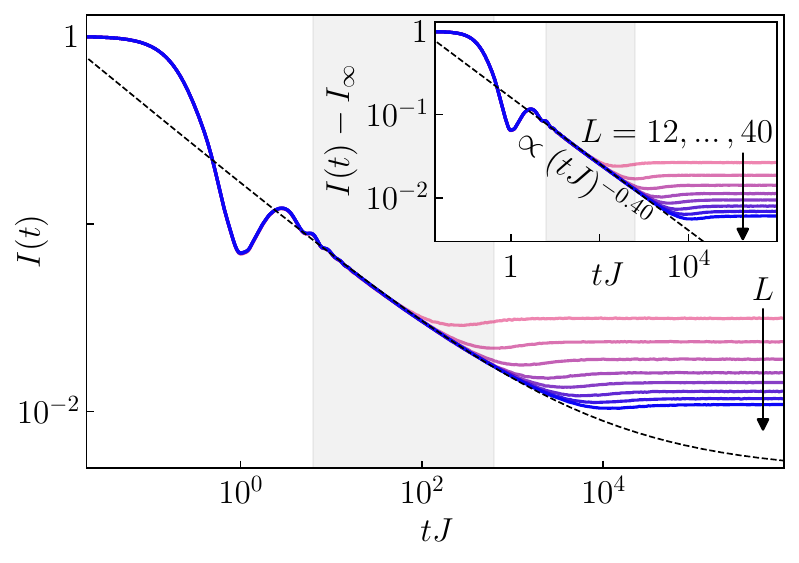}
\vspace{-0.2cm}
\caption{
Dynamics of the unscaled imbalance $I(t)$ [see Eq.~(\ref{eq:def_imbalance})], from the initial CDW state for the 3D Anderson model at the critical point $W_c/J=16.5$ and for system sizes $L = 12,16,20,24,28,32,36,40$.
The dashed line is a fit (within the shaded region at $L=40$) to the function $a (tJ)^{-\beta} + I_\infty$, where the nonzero infinite-time value in the thermodynamic limit $I_\infty$ is calculated from Eq.~(\ref{eq:I_scaling}).
Inset: The same as the main panel, but subtracted by $I_\infty$.
}
\label{fig11}
\end{figure}

\begin{figure}[!b]
\includegraphics[width=0.99\columnwidth]{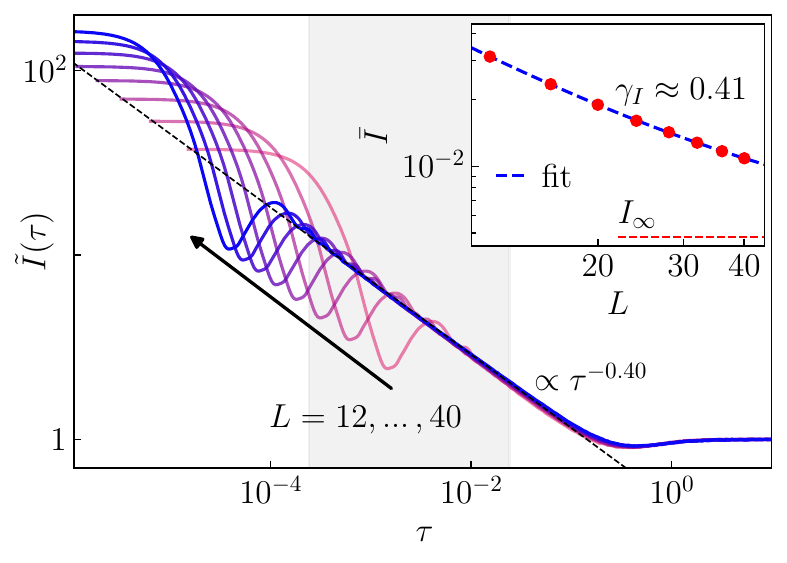}
\vspace{-0.2cm}
\caption{
Dynamics of the scaled imbalance $\tilde{I}(\tau)$ as a function of scaled time $\tau$ [see Eq.~(\ref{imbalance_rescaled})] from the initial CDW state for the 3D Anderson model at the critical point $W_c/J=16.5$ [the same data as in Fig.~\ref{fig11}].
The dashed line is a two-parameter fit (within the shaded region at $L=40$) to the function in Eq.~(\ref{def_i_tau_beta}).
Inset: Infinite-time values $\Bar{I}$ versus $L$ [circles] and the three-parameter fits to Eq.~(\ref{eq:I_scaling}) [dashed line]. 
The horizontal dotted line is the infinite-time value in the thermodynamic limit $I_\infty$.
}
\label{fig12}
\end{figure}

Nevertheless, as argued below, we indeed observe strong signatures of a scale-invariant dynamics in both $\tilde{n}(\tau)$ and $\tilde{I}(\tau)$. 
Our main observation is that the scale-invariant part of $\tilde{I}(\tau)$ can be described by the ansatz
\begin{align} \label{def_i_tau_beta}
    \tilde{I}(\tau) = a_I\,\tau^{-\beta_I} \;,
\end{align}
where $a_I$ and $\beta_I$ may or may not be identical to $a$ and $\beta$ from the approximations given in Eqs.~(\ref{eq:n_approximate}) and~(\ref{imbalance_approximate}).

{Next we test Eq.~(\ref{def_i_tau_beta}) numerically. We start with the 3D Anderson model and the initial ``chess'' CDW state discussed in Sec.~\ref{sec:case_study}.} 
Figure~\ref{fig11} shows the time evolution of the unscaled imbalance $I(t)$ (main panel), and the subtracted imbalance $I(t)-I_\infty$ (inset).
The latter exhibits a power-law decay that is scale invariant at mid times, but not at late times.
In Fig.~\ref{fig12}, we show the dynamics of $\tilde{I}(\tau)$, which exhibits a scale-invariant power-law decay at mid times, followed by a saturation to a scale-invariant plateau at late times.
As argued in Ref.~\cite{hopjan2023scaleinvariant}, the scale-invariant mid-time and late-time dynamics of the survival probability (i.e., the $d=0$ contribution) implies the scale-invariant mid-time dynamics of the unscaled survival probability, but not vice versa. 

The scale-invariant dynamics of the imbalance $\tilde{I}(\tau)$ from Fig.~\ref{fig12} is the main result of this paper.
It shows that Eq.~(\ref{def_i_tau_beta}) is indeed relevant for the description of the dynamics of the imbalance.
Before discussing its further properties, we make two comments. 
The {first} comment is that the scale invariance from 
Fig.~\ref{fig12} emerges only at the critical point.
In Appendix~\ref{sec:away}, we study {both ${I}(t)$ and} $\tilde{I}(\tau)$ in the 3D Anderson model below and above the critical point. {While the unscaled ${I}(t)$ exhibits arguably similar mid-time dynamics for all disorder strengths under investigation (compare Figs.~\ref{fig17a} and~\ref{fig17b} of Appendix~\ref{sec:away} to Fig.~\ref{fig11}),  we observe no scale invariance in the mid-time dynamics of $\tilde{I}(\tau)$} (see Fig.~\ref{fig15} of Appendix~\ref{sec:away}).
This justifies the notion of scale-invariant {\it critical} dynamics when referring to the results in Fig.~\ref{fig12}.

The {second} comment is that the result from Fig.~\ref{fig12} is not necessarily a consequence of the special CDW initial state, since other initial states, which also share the same eigenbasis with site occupation operators, show similar properties.
{As example we study the dynamics for other initial CDW states. In the 3D Anderson model, we construct the ``layered'' CDW state and ``striped'' CDW state, shown in Fig.~\ref{sketch}.
We note that for these initial states, the dynamics of site occupations $n_{\rm occ}(t)$, $n_{\rm unocc}(t)$ as well as the imbalance $I(t)$ is given by the same Eqs.$~$\eqref{eq:def_nt} and~\eqref{eq:def_imbalance} as for the chess CDW initial state studied in the main text.
However, when expressing $n_{\rm occ}(t)$ and $n_{\rm unocc}(t)$ as sums of single-particle transition probabilities, the particular contributions are different from those in Eqs.~\eqref{eq:def_densities_1} and~(\ref{eq:def_densities_2}). }

\begin{figure}[!t]
\includegraphics[width=0.95\columnwidth]{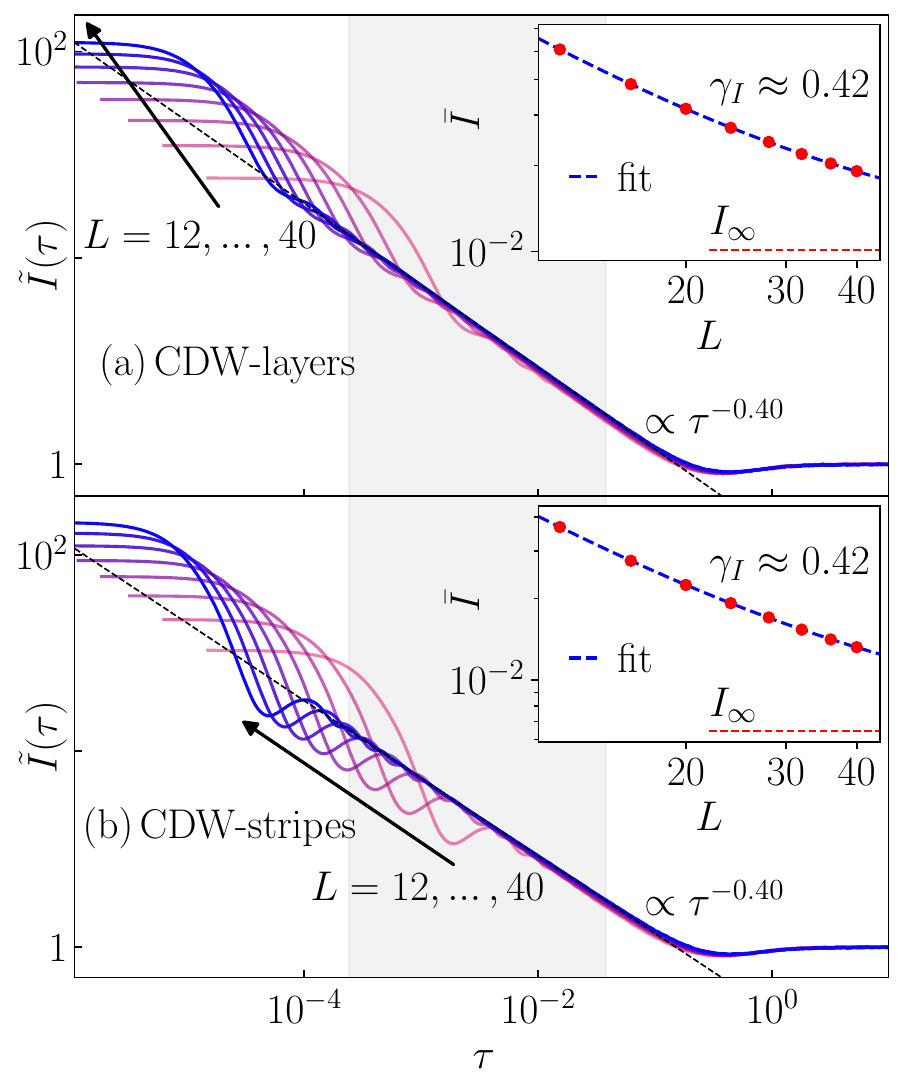}
\vspace{-0.2cm}
\caption{{
Dynamics of the scaled imbalance $\tilde{I}(\tau)$ as a function of scaled time $\tau$ [see Eq.~(\ref{imbalance_rescaled})] for the 3D Anderson model at the critical point $W_c/J=16.5$.
These results are analogous to the ones from Fig.~\ref{fig12}, but for different initial product states: (a) CDW-layers and (b) CDW-stripes (see the text and Fig.~\ref{sketch} for details). Insets: Infinite-time values $\Bar{I}$ versus $L$ (circles) and the three-parameter fits to Eq.~(\ref{eq:I_scaling}) (dashed line). 
The horizontal dotted line is the infinite-time value in the thermodynamic limit $I_\infty$.}
}
\label{fig16}
\end{figure}

{The dynamics of the scaled imbalance $\tilde{I}(\tau)$ for both additional CDW initial states is shown in Fig.$~$\ref{fig16} at the critical point of the 3D Anderson model.
We observe a scale-invariant power-law behavior of $\tilde{I}(\tau)$. Moreover, the power-law exponent $\beta_I$ is, in both cases, nearly identical to the one for the initial ``chess'' CDW state studied in Fig.~\ref{fig12} of the main text. }

\begin{figure}[!t]
\includegraphics[width=0.99\columnwidth]{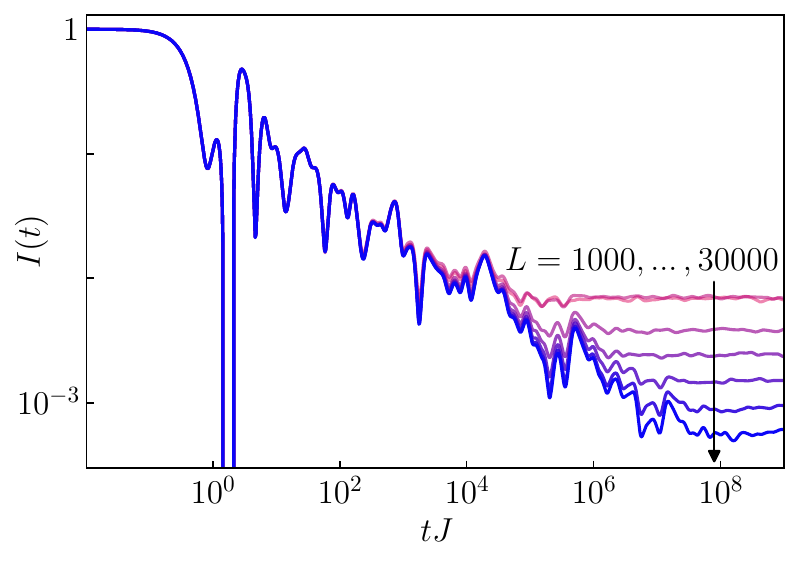}
\vspace{-0.2cm}
\caption{
Dynamics of the unscaled imbalance $I(t)$ [see Eq.~(\ref{eq:def_imbalance})] from the initial CDW state for the 1D Aubry–André model at the critical point $\lambda_c/J=2$ and for system sizes $L =$ 1000, 1250, 2500, 5000, 10$~$000, 20$~$000, 30 $~$000. To smooth the results, we used a Gaussian filter function with a standard deviation of 5 data points.  
}
\label{fig13}
\end{figure}

Next, we discuss the imbalance decay from the initial CDW state in the 1D Aubry–André model. In Fig.$~$\ref{fig13}, we plot $I(t)$ of the CDW state at the critical point of the 1D Aubry–André model.
The long-time dynamics may be interpreted as a power-law decay towards the infinite-time value, however, the temporal fluctuations in $I(t)$ are much stronger than those in the transition probabilities at a fixed distance $d$ shown in Fig.$~$\ref{fig4} for the same model.
The strong fluctuations may even yield negative values of $I(t)$  at short times (see Fig.~\ref{fig13}) in contrast to the 3D Anderson model, in which $I(t)$ is always positive (see Fig.~\ref{fig11}).

\begin{figure}[!t]
\includegraphics[width=\columnwidth]{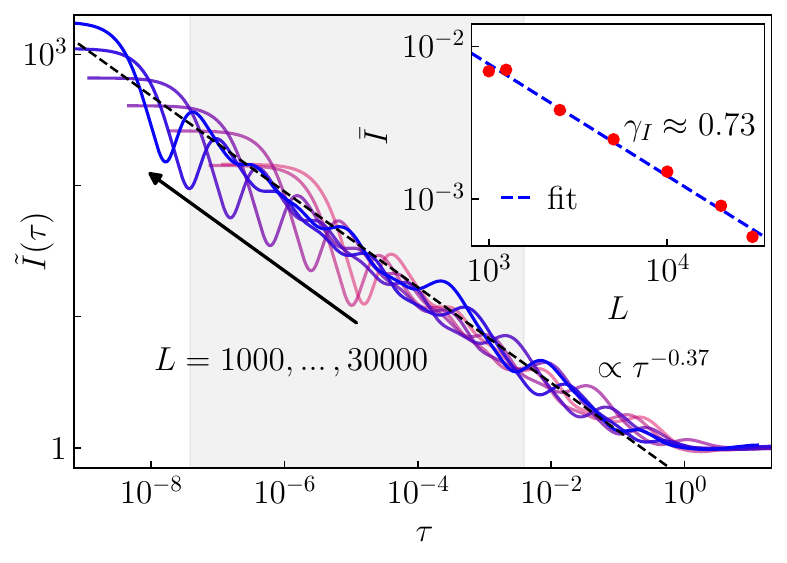}
\vspace{-0.2cm}
\caption{
Dynamics of the scaled imbalance $\tilde{I}(\tau)$ as a function of scaled time $\tau$ [see Eq.~(\ref{imbalance_rescaled}] from the initial CDW state for the 1D Aubry–André model at the critical point $\lambda_c/J=2$ (the same data as in Fig.~\ref{fig13}). To smoothen the results, we used a Gaussian filter function with a standard deviation of 20 data points. The dashed line is a two-parameter fit (within the shaded region at $L=$$~$30 000) to the function in Eq.~(\ref{def_i_tau_beta}).
Inset: Infinite-time value $\Bar{I}$ versus $L$ (circles) and the two-parameter fit to Eq.~(\ref{eq:I_scaling}) using $I_\infty=0$ (dashed line).
}
\label{fig14}
\end{figure}

In Fig.~\ref{fig14}, we plot the imbalance $\tilde{I}(\tau)$ in the 1D Aubry–André model.
To reduce the temporal fluctuations, we actually show the running averages of $\tilde{I}(\tau)$.
The resulting data are still not as smooth as those in the 3D Anderson model shown in Fig.~\ref{fig12}. Nevertheless, they appear to be consistent with the emergence of scale invariance in the critical dynamics of the imbalance.

\begin{figure}[!b]
\includegraphics[width=0.99\columnwidth]{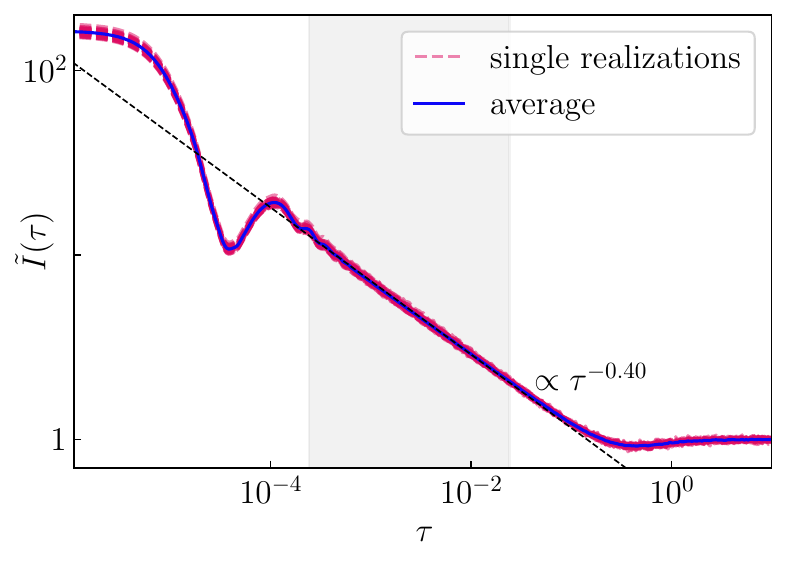}
\vspace{-0.2cm}
\caption{ {Dynamics of the scaled imbalance $\tilde I(t)$, from the initial CDW state for the 3D Anderson model at the critical point $W_c/J=16.5$ and for system size $L = 40$. Comparison of 50 independent Hamiltonian realizations (dashed lines) with the average over all Hamiltonian realizations (solid line).}
}
\label{fig22}
\end{figure}

To summarize, the main outcome of this section is  that the scale invariance indeed emerges in the critical dynamics of site occupations and the imbalance in the models under investigation.
This observation may justify the approximation made in Eq.~(\ref{eq:n_approximate}) to derive Eqs.~(\ref{eq:n_analytic}) and~(\ref{imbalance_approximate}).
It may appear, however, surprising that the scale invariance of $\tilde{I}(\tau)$ (see Figs.~\ref{fig12} and~\ref{fig14}) emerges at times that are comparable to those at the onset of the scale invariance of $p^{(d)}(\tau)$ at small $d$ but not large $d\gg 1$ (see Figs.~\ref{fig8} and~\ref{fig9}).
In other words, the scale invariance of observables may emerge despite the fact that the transition probabilities at large $d\gg 1$ do not yet reach the peak, and hence they have not yet entered the scale-invariant regime.

{Remarkably, we observe very similar dynamics for different Hamiltonian realizations. In Fig.~\ref{fig22} we compare the averaged scaled imbalance from Fig.~\ref{fig12} (solid line) to the unaveraged curves established from single Hamiltonian realizations of the 3D Anderson model (dashed lines). The latter exhibit an excellent agreement. Thus, it seems that the imbalance measure overcomes the need of averaging over Hamiltonian realization that is necessary, e.g., in the case of the survival probability measure. This opens a possibility to observe the critical dynamics even in a single Hamiltonian realization for a chosen system size $L$. We note, however, that even for one Hamiltonian realization, many repetitions of an experiment are still needed as the measurement is destructive.}

\subsection{Extraction of the fractal dimension}

A question beyond the mere emergence of scale invariance is whether one can extract some physically relevant information from the exponent of the power-law decay, e.g., from the decay exponent $\beta_I$ of the scaled imbalance in Eq.~(\ref{def_i_tau_beta}).
In the 3D Anderson model (see Fig.~\ref{fig12}) we obtain $\beta_I = 0.4$, which is very close to the decay exponents of the transition probabilities $\beta_d \in [0.40,0.45]$ (see Fig.~\ref{fig8}).
By contrast, we obtain $\beta_I=0.37$ in the 1D Aubry–André model, as shown in Fig.$~$\ref{fig14}.
This is different from the decay exponents of the transition probabilities $\beta_d=0.26$, which were reported in Fig.~\ref{fig9}.
While the difference between the values of $\beta_I$ and $\beta_d$ in the 1D Aubry–André model is not very large, we still find them less likely to agree in the thermodynamic limit, since the system sizes under investigation ($L\leq$30$~$000 lattice sites) are rather large.
This observation suggests that the concept of scale-invariant dynamics is more general and it emerges without the need for observables to decay with the same exponents as the survival and transition probabilities do.

It is still an interesting question to ask under which conditions  the exponent $\beta_I$ of the imbalance decay from Eq.~(\ref{def_i_tau_beta}) provides information about the fractal dimension $\gamma$ [cf.~Eq.~(\ref{eq:def_Pd_infty}) at $d=0$].
Studying the scale-invariant dynamics of survival probabilities, Refs.~\cite{hopjan2023, hopjan2023scaleinvariant} established a relationship between the fractal dimension $\gamma$ and the power-law exponent $\beta$ [more precisely, $\beta\equiv \beta_0$ in Eq.~(\ref{eq:scaling_ansatz})],
\begin{equation} \label{def_gamma_n_beta}
\gamma=n\beta\;,
\end{equation}
where $n$ determines the scaling of the typical Heisenberg time with the system size at the critical point, $t_H^{\rm typ} = D^n$ ($n\approx 1$ in the 3D Anderson model and $n\approx 2$ in the 1D Aubry–André model~\cite{hopjan2023}).
For the imbalance, one can define the corresponding exponent $\gamma_I$ as
\begin{equation} \label{eq:I_scaling}
\overline{I}=c_I\,D^{-\gamma_I}+I_\infty \;, 
\end{equation}
which is numerically extracted in the insets of Figs.~\ref{fig12} and~\ref{fig14} at the critical points of the 3D Anderson and 1D Aubry-André models, respectively.

At this point one can ask two questions. First, may one expect the analog of Eq.~(\ref{def_gamma_n_beta}) for imbalance, i.e., $\gamma_I = n \beta_I$?
Second, if this is the case, under what circumstances will one observe $\gamma_I = \gamma$?
These are exciting questions for which, however, we are not able yet to provide a conclusive answer.

Nevertheless, a numerical analysis of $\beta_I$ and $\gamma_I$ at the critical point may provide some useful insights into these questions. 
First, the results from Fig.~\ref{fig12} suggest that $\beta_I\approx0.4$ and $\gamma_I\approx0.4$ in the 3D Anderson model
while the those from Fig.~\ref{fig14} suggest that $\beta_I\approx0.37$ and $\gamma_I\approx0.73$ the 1D Aubry–André model.
This is indeed consistent with Eq.~(\ref{def_gamma_n_beta}) and hence we conjecture that a relationship
\begin{equation} \label{def_gamma_n_beta_I}
\gamma_I = n \beta_I\;
\end{equation}
applies to both models.
The validity of Eq.~(\ref{def_gamma_n_beta_I}) can also be interpreted as a consequence of the scale-invariant power-law behavior of the scaled imbalance $\tilde I(\tau)$ using similar arguments as those for the survival probability in Ref.~\cite{hopjan2023}.
As argued in Ref.~\cite{hopjan2023scaleinvariant}, these properties also give rise to the scale-invariant mid-time dynamics of the subtracted but unscaled imbalance $I(t)-I_\infty$, which is indeed observed in the dynamics of the imbalance at the critical point in both models (see Figs.~\ref{fig11} and~\ref{fig13}).

Remarkably, in the case of the 3D Anderson model, we further observe that $\gamma_I$ is quantitatively similar to the fractal exponent, i.e., $\gamma_I \approx \gamma$, and thus the decay of the imbalance also appears to provide information about the fractal dimension $\gamma$. 
Some further insights as to why this is the case for the the 3D Anderson model but not for the 1D Aubry-André model, and to which degree $\gamma_I$ reflects the presence of multifractality at criticality, are given in Appendix~\ref{sec:imbalancedecay}.

\section{Conclusions} \label{sec:conclusions}

This paper explores to what extent one may observe scale-invariant behavior in the nonequilibrium quantum dynamics of one-body observables at criticality.
We considered quadratic Hamiltonians and a class of observables that are diagonal in the single-particle eigenbasis of the Hamiltonian before the quench.
Namely, the dynamics of these observables expressed in many-body states equal the sum of transition probabilities of the initially occupied single-particle states to other states.
Consequently, the quest for  scale invariance of observables translates to the study of the scale invariance of the transition probabilities.

We numerically studied the transition probabilities in two quadratic fermionic models, the 3D Anderson model and the 1D Aubry–André model, focusing on the CDW initial states.
Considering the initial single-particle states as site-localized states, we observe scale invariance in the dynamics of transition probabilities at criticality that shares strong similarities with the previously studied scale invariance of the survival probabilities~\cite{hopjan2023, hopjan2023scaleinvariant}.
Both models exhibit a temporal power-law decay with an exponent $\beta$ that appears to be very similar for all transition probabilities, as well as for the survival probability.
We conjecture that all transition probabilities become scale invariant in the thermodynamic limit.
Our numerical results strongly support this conjecture for the 3D Anderson model and to a large degree also in the 1D Aubry–André model. However, the latter exhibits stronger finite-size fluctuations.

Having established the scale invariance in the dynamics of transition probabilities as a stepping stone for the scale invariance of observables, we gave heuristic arguments for the scale invariance of the observables such as the scaled site occupations $\tilde n(\tau)$ and imbalance $\tilde{I}(\tau)$. We then numerically showed that $\tilde{I}(\tau)$ becomes scale invariant even for system sizes at which some of the transition probabilities $p^{(d)}(\tau)$, at  distances $d\gg 1$, are not yet in the scale-invariant regime. Intriguingly, the notion of scale invariance appears to be a general principle for the critical dynamics. This principle opens possibilities to detect fingerprints of criticality in experimentally relevant observables at relatively short time scale, {and calls for further studies of interacting models when interpreting them as single-particle models on a Fock-space graph}.

At the same time, such principle leads to the generalization of the relationship in Eq.~(\ref{def_gamma_n_beta}) between the fractal dimension $\gamma$ and the exponent $\beta$ of the temporal power-law decay (see Refs.~\cite{hopjan2023, hopjan2023scaleinvariant}) to a similar relationship for the imbalance, i.e., $\gamma_I=n\beta_I$ from Eq.~(\ref{def_gamma_n_beta_I}).
Remarkably, for the 3D Anderson model in which the imbalance can be well approximated by the contributions of transition probabilities at small distances, the resulting decay coefficient $\gamma_I$ is close to the fractal dimension $\gamma$. 
Our results hence open a possibility to extract the fractal dimension $\gamma$ from the temporal decay exponent $\beta_I$ of experimentally relevant observables, which calls for further insights from both numerical and analytical sides.

Another important question is the role of the initial single-particle states for the identification of critical dynamics.
From the perspective of the survival probability, Ref.~\cite{hopjan2023scaleinvariant} suggested that the scale invariance at criticality may also be observed for initial single-particle states different from site-localized states, which do not exhibit a power-law decay.
The generalization of these results to arbitrary one-body observables should be explored in future work.\\

Research data are available as ancillary files ~\cite{jiricek2024criticalquantumdynamicsobservables}.

\acknowledgements 
This work is funded by the Deutsche Forschungsgemeinschaft (DFG, German Research Foundation) – 499180199, 436382789, 493420525,  via FOR 5522  and large-equipment grants (GOEGrid cluster).
This research was supported in part by the National Science Foundation under Grant No. NSF PHY-1748958. F.H.-M. is grateful for the hospitality at KITP, UC Santa Barbara, where part of this work was performed.
We acknowledge support from the Slovenian Research and Innovation Agency (ARIS), Research core funding Grants No.~P1-0044, N1-0273, J1-50005 and N1-0369 (M.H. and L.V.).
We gratefully acknowledge the High Performance Computing Research Infrastructure Eastern Region (HCP RIVR) consortium~\cite{vega1} and
European High Performance Computing Joint Undertaking (EuroHPC JU)~\cite{vega2}  for funding this research by providing computing resources of the
HPC system Vega at the Institute of Information sciences~\cite{vega3}.
This work used the Scientific Compute Cluster at GWDG, the joint data center of Max Planck Society for the Advancement of Science (MPG) and University of G\"ottingen.
\appendix

\section{Imbalance away from criticality} \label{sec:away}

In the main text, we have focused on the dynamics at the critical point of both models under investigation.
Here we study whether the notion of scale-invariant dynamics of observables, established in Sec.~\ref{sec:observables}, also extends to the regimes away from the critical point.

\begin{figure}[!t]
\includegraphics[width=0.99\columnwidth]{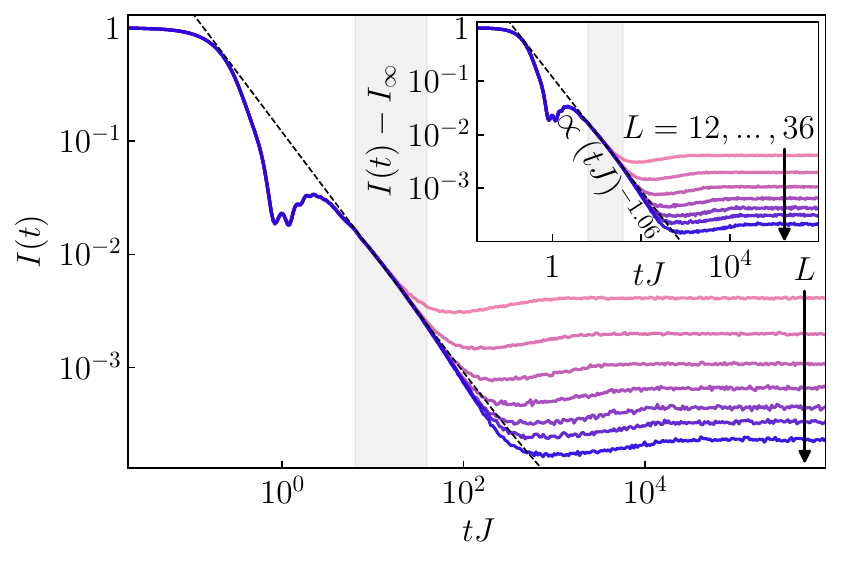}
\vspace{-0.2cm}
\caption{{
Dynamics of the unscaled imbalance $I(t)$ [see Eq.~(\ref{eq:def_imbalance})] from the initial CDW state for the 3D Anderson model in the chaotic regime $W/J=10$ and for system sizes $L = 12,16,20,24,28,32,36$.
The dashed line is a fit (within the shaded region at $L=36$) to the function $a (tJ)^{-\beta} + I_\infty$, where the nonzero infinite-time value in the thermodynamic limit $I_\infty$ is calculated from Eq.~(\ref{eq:I_scaling}).
Inset: The same as the main panel, but subtracted by $I_\infty$.}
}
\label{fig17a}
\end{figure}

\begin{figure}[!b]
\includegraphics[width=0.99\columnwidth]{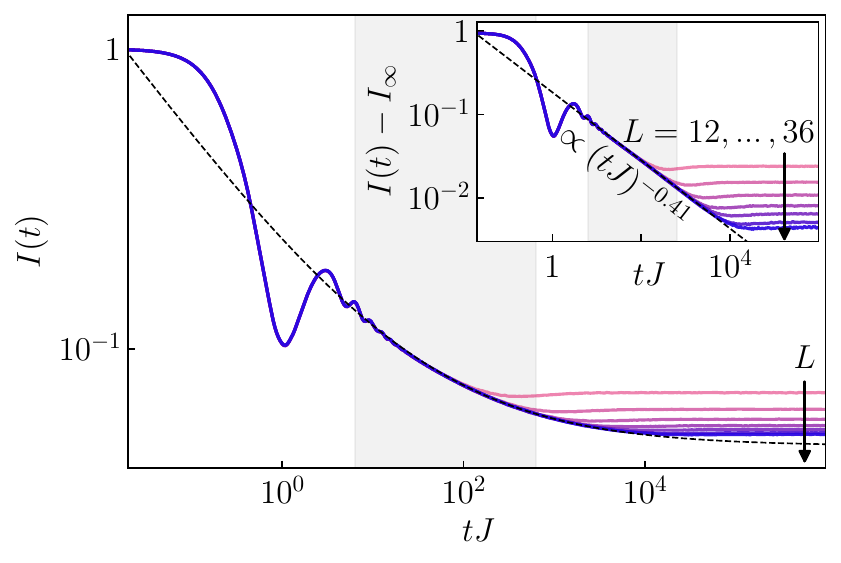}
\vspace{-0.2cm}
\caption{{
Dynamics of the unscaled imbalance $I(t)$ [see Eq.~(\ref{eq:def_imbalance})] from the initial CDW state for the 3D Anderson model in the localized regime $W/J=20$ and for system sizes $L = 12,16,20,24,28,32,36$.
The dashed line is a fit (within the shaded region at $L=36$) to the function $a (tJ)^{-\beta} + I_\infty$, where the nonzero infinite-time value in the thermodynamic limit $I_\infty$ is calculated from Eq.~(\ref{eq:I_scaling}).
Inset: The same as the main panel, but subtracted by $I_\infty$.}
}
\label{fig17b}
\end{figure}

{
Figure~\ref{fig17a} shows results for the time evolution of the unscaled imbalance $I(t)$ (main panel) and the subtracted imbalance $I(t)-I_\infty$ (inset), for the 3D Anderson model at $W/J=10$ that corresponds to the regime of single-particle quantum chaos.
We observe a power-law like decay of $I(t)$ that is scale invariant at mid times, but not at late times. The slope of the decay appears to get steeper at later times with the increasing system size, and the effect of the mobility edges is almost invisible.  Figure~\ref{fig17b} shows the identical quantities at $W/J=20$ that corresponds to the localized regime.
We observe a decay of $I(t)$ towards the value $I_\infty$ that is again scale invariant at mid times, but not at late times. The inset of Fig.~\ref{fig17b} reveals that the decay towards a constant follows a power-law.
}

{
In all the cases under consideration (i.e., at the transition and in the chaotic and localized regimes), we observe that the dynamics of $I(t)$ are scale-invariant at short and mid times. A similar behavior is observed for the survival probability $P^{(0)}(t)$ (not shown). 
We note that the short-time behavior of the latter is governed by the width of the local density of states (LDOS)~\cite{Schiulaz_19,das2024proposal}.
In the case of short-range models, which are studied in this work, the width of the LDOS  does not increase with the system size since the number of neighboring lattice sites is kept constant, and hence the short-time behavior is scale invariant.
We believe that the same reasoning applies to the imbalance $I(t)$.}

\begin{figure}[!t]
\includegraphics[width=0.95\columnwidth]{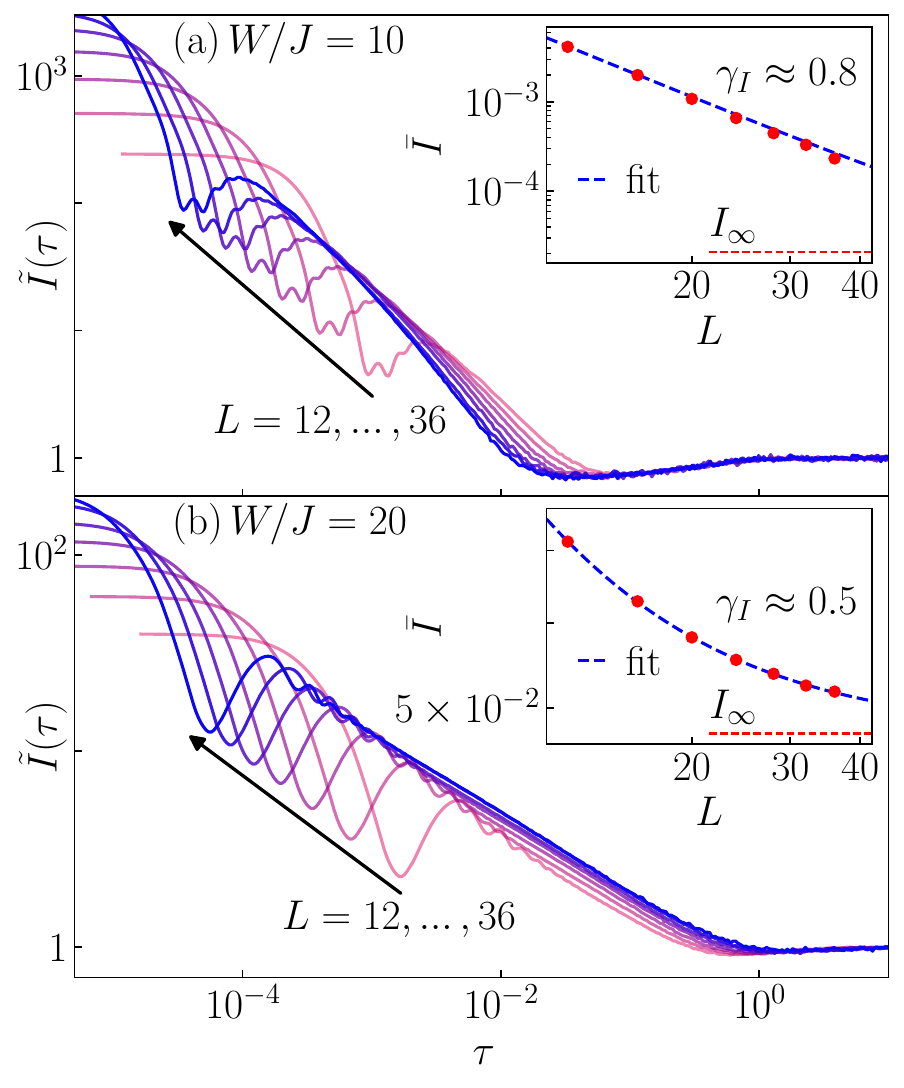}
\vspace{-0.2cm}
\caption{{
Dynamics of the scaled imbalance $\tilde{I}(\tau)$ as a function of scaled time $\tau$ [see Eq.~(\ref{imbalance_rescaled})] from the initial CDW state for the 3D Anderson model.
The results are analogous to those from Fig.~\ref{fig12}, but shown (a) in the chaotic regime $W/J=10$ and (b) in the localized regime $W/J=20$.
Insets: Infinite-time values $\Bar{I}$ versus $L$ (circles) and the three-parameter fits to Eq.~(\ref{eq:I_scaling}) (dashed line). 
The horizontal dotted line is the infinite-time value in the thermodynamic limit $I_\infty$.}}
\label{fig15}
\end{figure}

{
Focusing on longer time scales, we first note that, as argued in Ref.~\cite{hopjan2023scaleinvariant}, the scale-invariant mid-time dynamics of the unscaled survival probability $P^{(0)}(t)$ does not imply scale invariance of the mid-time and late-time dynamics of the scaled survival probability $p^{(0)}(\tau)$.
Here we argue that the same applies to the imbalance $I(t)$, i.e., the scale-invariant mid-time dynamics of the the unscaled imbalance $I(t)$ shown in Figs.~\ref{fig17a} and~\ref{fig17b} (corresponding to $W/J=10$ and 20, respectively) does not imply scale invariance of the mid-time and late-time dynamics of the {\it scaled} imbalance $\tilde{I}(\tau)$.
The corresponding results for the latter are shown in Figs.~\ref{fig15}(a) and~\ref{fig15}(b) at $W/J=10$ and 20, respectively, indicating the absence of scale invariance.
While at $W/J=10$, the decay as a function of $\tau$ becomes faster with increasing $L$, indicating relaxation at times shorter than the Heisenberg time, at $W/J=20$, the decay becomes slower with increasing $L$. 
}

{
We observe that a similar absence of scale invariance is also present in the 1D Aubry–André model away from the criticality (not shown). The results are similar to the absence of scale invariance in the dynamics of scaled survival probabilities away from the criticality, studied in Ref.~\cite{hopjan2023}. }

\begin{figure}[!t]
\includegraphics[width=0.99\columnwidth]{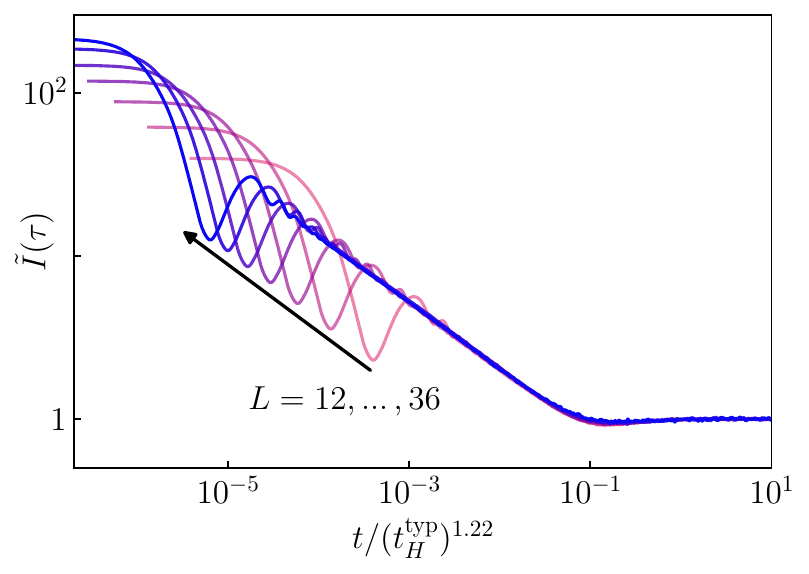}
\vspace{-0.2cm}
\caption{{Dynamics of the scaled imbalance $\tilde{I}(\tau)$ as a function of scaled time $\tau = t/(t_H^{\rm typ})^{a}$, with a modified scaling exponent $a$ from the initial CDW state for the 3D Anderson model [the same data as in Fig.~\ref{fig15}(b)].}}
\label{fig17c}
\end{figure}

{
We interpret the emergence of scale invariance at the critical point a consequence of the property that only exists at the critical point, namely, the longest relaxation time (also known as the Thouless time) equals the Heisenberg time~\cite{sierant_delande_20, suntajs_prosen_21}.
Still, it is an interesting question whether there exist other forms of scaling collapses, in particular in the localized phase.
To address this question, one should go beyond the simple rescaling of time with the Heisenberg time, $\tau=t/t_H^{\rm typ}$.
For example, one could introduce a parameter $a$ that is different from 1, such that a generalized time is expressed as $t/(t_H^{\rm typ})^{a}$ (in dimensionless units).
In Fig.~\ref{fig17c} we show that an excellent scaling collapse can be obtained in the localized phase at $W/J=20$ for $a=1.22$. 
The property $a>1$ is consistent with the drift of the Thouless time towards values larger than the Heisenberg time. 
}

\section{Fractal dimension versus imbalance decay}
\label{sec:imbalancedecay}

We here discuss under what circumstances the exponent $\gamma_I$ from Eq.~(\ref{eq:I_scaling}) may equal the fractal dimension $\gamma$.
Interestingly, for the 3D Anderson model, the temporal decay exponent $\beta_I$ from Eq.~(\ref{def_i_tau_beta}) is very similar to the analogous decay exponent $\beta$ of survival and transition probabilities, i.e., $\beta_I\approx\beta$, as seen from Figs.~\ref{fig8}(a) and~\ref{fig12}.
In this case, Eq.~(\ref{def_gamma_n_beta_I}) then implies $\gamma_I\approx\gamma$. To understand this behavior it is instructive to consider the following cumulative function,
\begin{equation} \label{eq:Cum}
    \overline{C}(d^{*})= \sum_{d=0}^{d^{*}} (-1)^{d}\, \overline{P}^{(d)} \;,
\end{equation}
which for the CDW initial state considered in the main text gives $\overline{C}(d_{\rm max})=\bar{I}$ at the maximal distance $d^{*}=d_{\rm max}$.

\begin{figure}[!t]
\includegraphics[width=1.0\columnwidth]{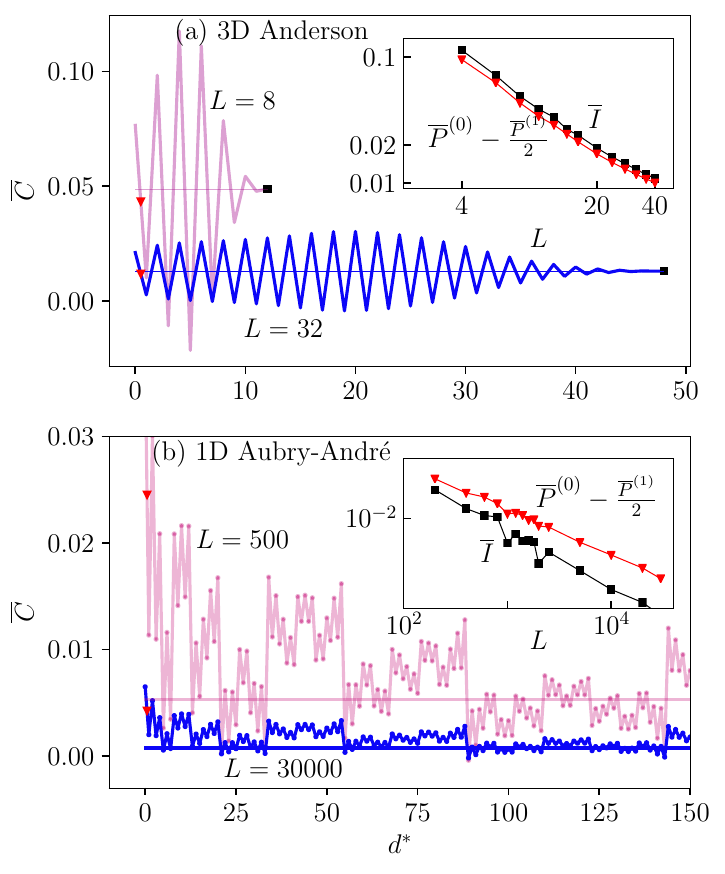}
\vspace{-0.4cm}
\caption{Cumulative function $\overline{C}(d^{*})$ from Eq.~(\ref{eq:Cum}) as a function of the radius $d^{*}$ for the CDW initial state at the critical points of (a) the 3D Anderson model and (b) the 1D Aubry-André model.
The system sizes $L$ are listed in the figures.
The value of imbalance  $\overline{C}(d_{\rm max})=\overline{I}$ is denoted by the horizontal lines and black squares. The approximation~(\ref{eq:I_approx}) for the imbalance, $\overline{P}^{(0)}-\overline{P}^{(1)}/2$, is displayed with  red triangles.
The insets show the system-size dependence of $\overline{I}$ and $\overline{P}^{(0)}-\overline{P}^{(1)}/2$.}
\label{fig21}
\end{figure}

Examples of the cumulative function $\overline{C}(d^{*})$ are plotted in the main panels of Figs.~\ref{fig21}(a) and~\ref{fig21}(b) for the 3D Anderson model and the 1D Aubry–André model, respectively.
One observes that the cumulative function for the 3D Anderson model displays oscillatory behavior around the value of $\overline{I}$ [see the main panel of Fig.~\ref{fig21}(a)], while for the 1D Aubry–André model, the behavior is rather erratic [see  the main panel of Fig.~\ref{fig21}(b)].
In the case of oscillatory behavior observed in Fig.~\ref{fig21}(a), we assume that the value of the imbalance can be well approximated by the average value of the lowest-order cumulants.
Specifically, we approximate $\overline{I}$ as
\begin{equation} \label{eq:I_approx}
\overline{I}\approx [\overline{C}(0)+\overline{C}(1)]/2=\overline{P}^{(0)}-\overline{P}^{(1)}/2\;,
\end{equation}
where the left-hand side~of Eq.~(\ref{eq:I_approx}) is plotted as horizontal lines in the main panel of Fig.~\ref{fig21}(a) and the right-hand side~of Eq.~(\ref{eq:I_approx}) as red triangles.
The system-size dependence of $\overline{I}$ and $\overline{P}^{(0)}-\overline{P}^{(1)}/2$ is then compared in the inset of Fig.~\ref{fig21}(a).
We observe that this approximation holds for all system sizes considered and both quantities follow a nearly identical decay. As $\overline{P}^{(0)}$ and $\overline{P}^{(1)}$ both decay with the exponent $\gamma_d\approx\gamma$, the decay exponent for imbalance is $\gamma_I\approx\gamma$.

For the 1D Aubry–André model, however, the approximation from Eq.~(\ref{eq:I_approx}) does not work [see the inset of Fig.~\ref{fig21}(b)]. Here, the imbalance $\overline{I}$ decays faster than $\overline{P}^{(0)}-\overline{P}^{(1)}/2$.
The reason is that the higher-order terms $\overline{P}^{(d)}$ do not only cause oscillations around the mean value $\overline{I}$ but rather give rise to erratic jumps.
We hence conclude that at the critical point of the 3D Anderson model, $\gamma_I  \approx \gamma$, while this is not the case for the 1D Aubry–André model.

Even though the decay exponent $\gamma_I$ does not necessarily have the same value as $\gamma$, it still displays a nonzero value smaller than the one at the critical point.
This property can be interpreted as a (multi)fractal character of $\gamma_I$.
A necessary condition for this interpretation to be reasonable is that on the delocalized (chaotic) side of the eigenstate transition, one gets $\gamma_I\approx1$.

\begin{figure}[!t]
\includegraphics[width=1.0\columnwidth]{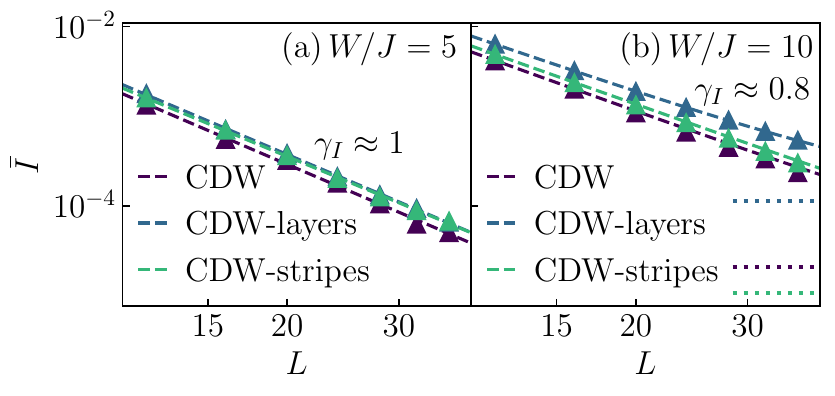}
\vspace{-0.2cm}
\caption{ 
The infinite-time value $\overline{I}$ of the imbalance for the 3D Anderson model as a function of system size $L$ below the critical point, at (a) $W/J=5$ and at (b) $W/J=10$, for the three initial CDW states studied in this work.
The decay is fitted using Eq.~(\ref{eq:I_scaling}), for which the saturation value is $I_\infty=0$ in (a) and $I_\infty>0$ in (b).
The fits are shown by dashed lines and the dotted horizontal lines in (b) represent $I_\infty$.
}
\label{fig20}
\end{figure}

We first checked (not shown) that for the GOE Hamiltonian matrices, the decay exponent indeed equals $\gamma^{\rm GOE}_I=1$.
Then, focusing on the chaotic regime of the 3D Anderson model (i.e., below the critical point), in Fig.~\ref{fig20}, we plot the decay of the imbalance $\overline{I}$ as a function of $L$.
We consider all the initial CDW states discussed in the main text.
We first note that, at fixed $W$, all initial states yield a similar decay exponent $\gamma_I$. Deep in the chaotic regime at $W/J=5$ [see Fig.~\ref{fig20}(a)] the decay of $\overline{I}$ is governed by the exponent $\gamma_I\approx1$.
This value is equal to the value of the decay exponent $\gamma\approx1$ of the survival probability $\overline{P}$ for both the 3D Anderson model at $W/J=5$ (not shown) and the GOE matrices~\cite{Schiulaz_19, hopjan2023scaleinvariant}, 
in agreement with the expected value of the fractal dimension in chaotic systems~\cite{Schiulaz_19, hopjan2023scaleinvariant}.
At a higher disorder strength such as $W/J=10$ [see Fig.~\ref{fig20}(b)], the decay exponent is $\gamma_I\approx0.8$, i.e., it  differs from the GOE limit.
We note that at the same disorder strength, $W/J=10$, the survival probability exhibits a similar decay exponent $\gamma\approx0.8$~\cite{hopjan2023}. 
The deviations of $\gamma_I$ and $\gamma$ at $W/J=10$ from the GOE limit are likely due to finite-size effects, and hence they should be considered as functions of system size before they enter the asymptotic regime. 
Indeed, a closer inspection shows that the slope of $\overline{I}(L)$ changes with the system size $L$ in Fig.~\ref{fig20}(b), and for even larger systems, they likely approach the GOE limit. 
Oppositely, according to the general scaling arguments~\cite{Abrahams79,Rodriguez09,Rodriguez10}, such a flow of the fractal dimension $\gamma$ is not expected at the critical point, and hence it is expected to stay smaller than one in the thermodynamic limit.
The same property is expected for $\gamma_I$.

For the 1D Aubry–André model on the delocalized side, one observes $\gamma \approx1$~\cite{hopjan2023} and we checked (not shown) that it also yields $\gamma_I\approx1$. 
It is expected that in the thermodynamic limit, $\gamma$ jumps from  one to a nonzero value smaller than one at the critical point.
Despite observing $\gamma_I\neq\gamma$ for the 1D Aubry–André model, our numerical results suggest that  $\gamma_I$ also jumps from  one to a nonzero value smaller than one at the critical point.

\section{Averaging of transition probabilities}\label{sec:averaging}

On the one hand, in Sec.$~$\ref{sec:case_study}, we introduced the transition probabilities $P_{\Psi_0}^{\,(d)}(t)$ in Eqs.~(\ref{def_Ppsi_d_1}) and~(\ref{def_Ppsi_d_2}), which were averaged over the initially occupied and unoccupied sites in the CDW state, respectively.
On the other hand, in Eq.~(\ref{eq:def_Pd}) in Sec.$~$\ref{sec:transition_prob}, we introduced the transition probabilities $P^{(d)}(t)$ as an average over all sites. 
In the first case, the averaging is carried out over $\frac{1}{2}$ of all sites and hence it depends on the particular structure of the initial state, while in the second case, the averaging is independent of the details of the initial state.

\begin{figure}[!t]
\includegraphics[width=0.99\columnwidth]{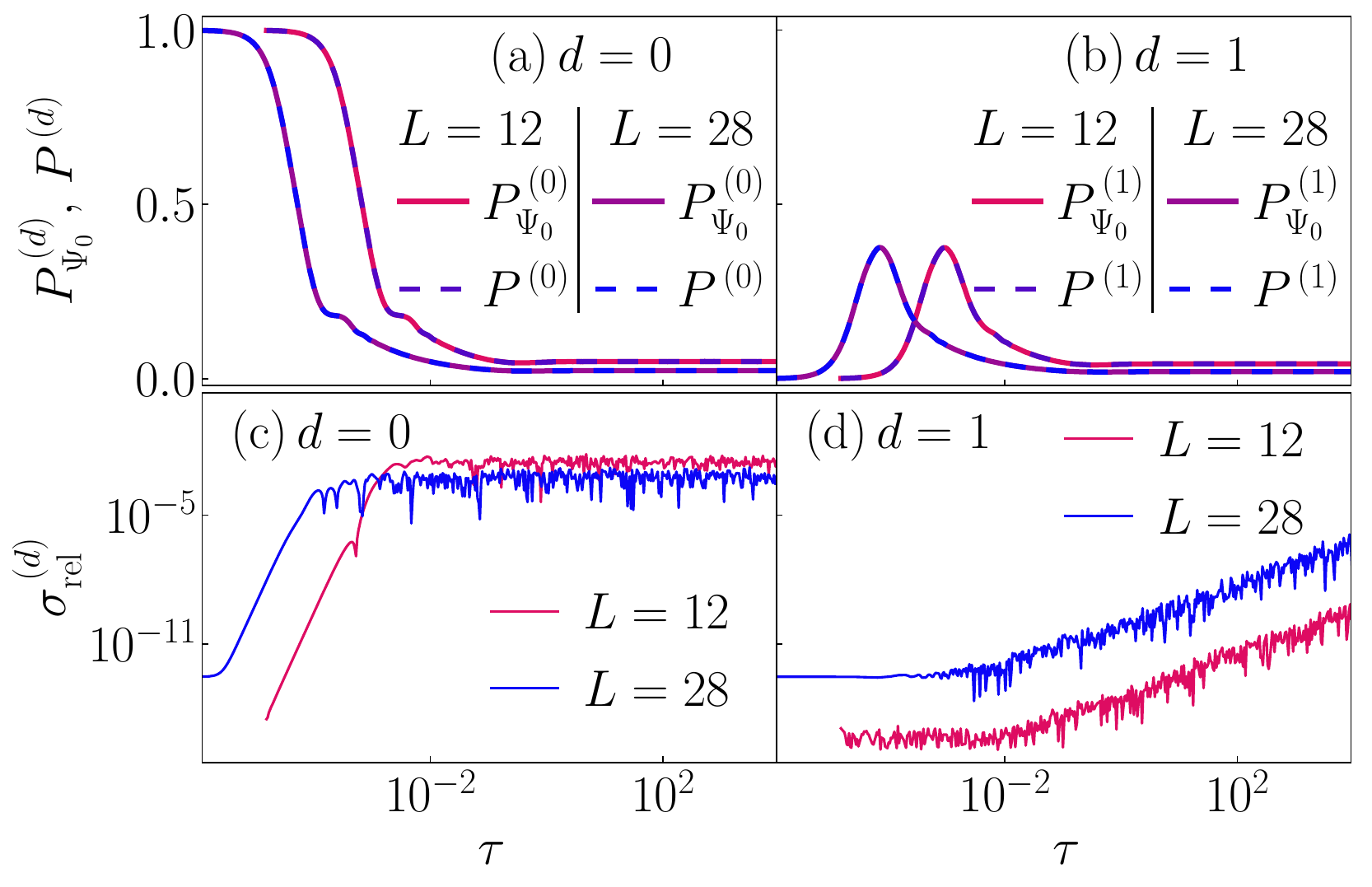}
\vspace{-0.2cm}
\caption{Comparison between the transition probabilities $P_{\Psi_0}^{\,(d)}(\tau)$ and $P^{\,(d)}(\tau)$ (see text for details) for the initial CDW state for the 3D Anderson model at $W_c/J=16.5$ as a function of scaled time $\tau$. 
(a) and (b): $P_{\Psi_0}^{\,(d)}(\tau)$ versus $P^{\,(d)}(\tau)$ at $d=0$ and 1, respectively, for two system sizes $L=12$ and 28.
The results are obtained from averages over 200 Hamiltonian realizations. 
(c) and (d): The corresponding relative deviation $\sigma^{\,(d)}_\mathrm{rel}(\tau)$ from Eq.~(\ref{eq:errors_P}) for two system sizes $L=12$ and 28.}
\label{fig17}
\end{figure}

\begin{figure}[!b]
\includegraphics[width=1.0\columnwidth]{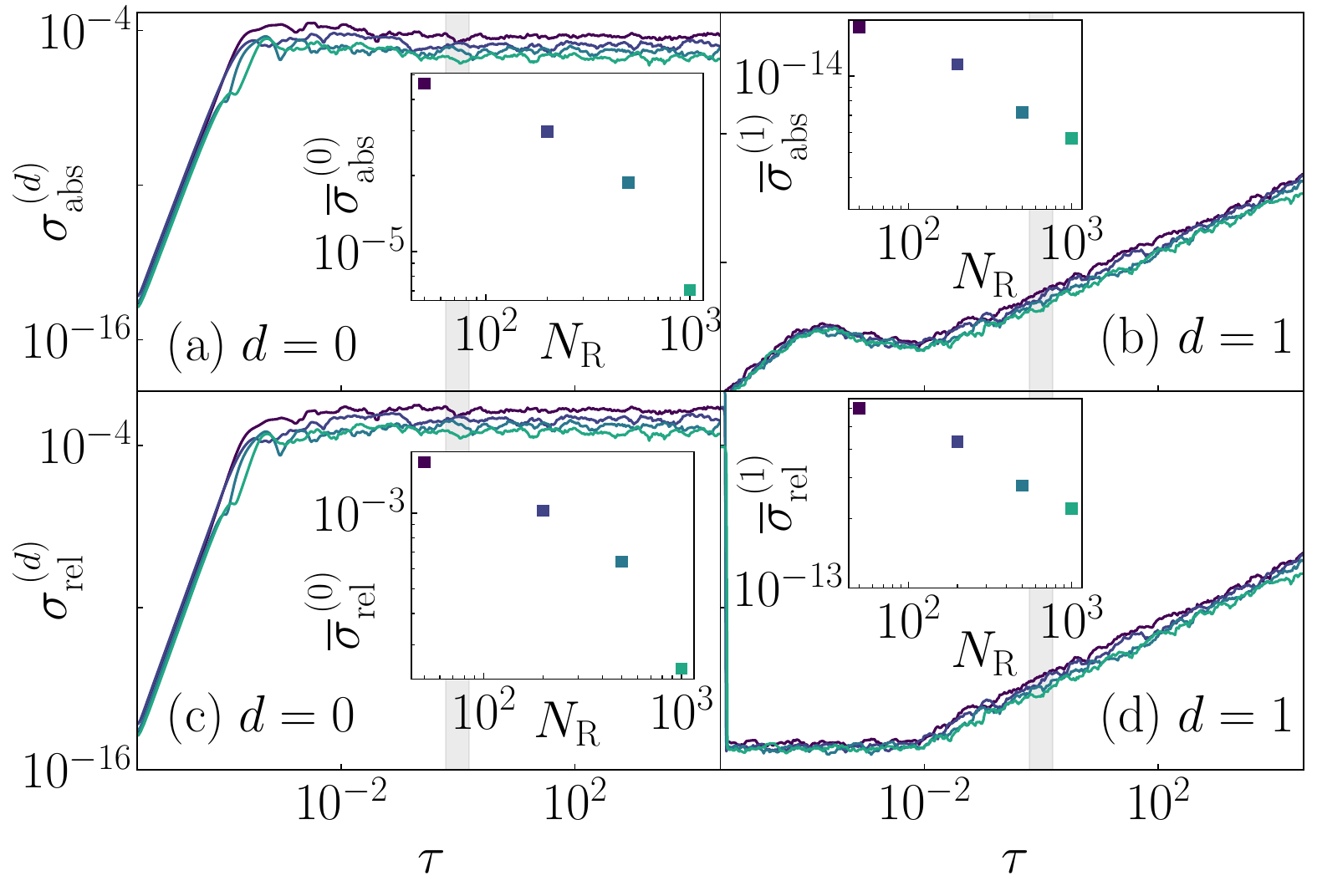}
\vspace{-0.2cm}
\caption{
Properties of the absolute and relative deviations $\sigma^{\,(d)}_\mathrm{abs}$ and $\sigma^{\,(d)}_\mathrm{rel}$, respectively, from Eq.~(\ref{eq:errors_P}) in the 3D Anderson model at $W_c/J=16.5$ and $L=20$.
(a), (b):
The absolute deviations $\sigma^{\,(d)}_\mathrm{abs}(\tau)$ at $d=0$ and 1, respectively, as a function of scaled time $\tau$.
(c), (d):
The relative deviations $\sigma^{\,(d)}_\mathrm{rel}(\tau)$ at $d=0$ and 1, respectively, as a function of scaled time $\tau$.
Different curves correspond to different numbers of Hamiltonian realizations $N_\text{R}$ and the color scale is given in the insets.
The insets show the mean deviations $\overline{\sigma}^{\,(d)}_\mathrm{abs}$ and $\overline{\sigma}^{\,(d)}_\mathrm{rel}$ in a small window, indicated by the shaded area in (a)--(d), around the typical Heisenberg time corresponding to $\tau\approx 1$, as functions of $N_\text{R}$.}
\label{fig18}
\end{figure}

The main goal of this section is to show that the two types of averaging described above do not yield significant differences.
We define the absolute and the relative deviation,
\begin{equation} \label{eq:errors_P}
    \sigma^{\,(d)}_\mathrm{abs}=|P^{\,(d)}-P_{\Psi_0}^{\,(d)}|\;,\;\;\;
    \sigma^{\,(d)}_\mathrm{rel}=\sigma^{\,(d)}_\mathrm{abs} / P^{\,(d)} \;,
\end{equation}
respectively, and we study their properties at $d=0, 1$ in Figs.$~$\ref{fig17} and~\ref{fig18} for the initial CDW state from Eq.~(\ref{def_cdw_state}).

We first compare the transition probabilities $P_{\Psi_0}^{\,(d)}(\tau)$, $P^{\,(d)}(\tau)$ at $d=0$ and 1 in Figs.~\ref{fig17}(a) and~\ref{fig17}(b), respectively.
They are virtually indistinguishable.
Their relative deviations $\sigma^{\,(d)}_\mathrm{rel}$ from Eq.~(\ref{eq:errors_P}), shown in Figs.~\ref{fig17}(c) and~\ref{fig17}(d), are very small.
They are of the order $10^{-4}$ at $d=0$ and $\lesssim 10^{-10}$ at $d=1$ and $\tau<1$.

The results from Figs.~\ref{fig17}(c) and~\ref{fig17}(d) show the dependence on the system size $L$ at a fixed number of Hamiltonian realizations $N_R=200$, while the results in Fig.~\ref{fig18} show the dependence on $N_R$ at a fixed system size $L=20$.
At $d=0$, where the deviations are the largest [see Figs.~\ref{fig17}(c) and~\ref{fig18}(c)] the deviations decrease both with increasing $L$ and $N_R$.
At $d=1$, where the deviations are practically negligible [see Figs.~\ref{fig17}(d) and~\ref{fig18}(d)] the deviations increase with $L$ but decrease with $N_R$.
Nevertheless, we conclude that the differences between the transition probabilities $P_{\Psi_0}^{\,(d)}(\tau)$ and $P^{\,(d)}(\tau)$ are very small and do not affect the  results of this paper, in particular, the observation of scale invariance of $P^{\,(d)}(\tau)$ and their relationship with the scale invariance of observables.

\section{Details about disorder averaging}
\label{sec:disorder_statistics_3D_Anderson}

\renewcommand{\arraystretch}{1.1}

\begin{table}[H]
\centering
\begin{tabular}{|c|c|c|c|c|}
\hline
$N_\text{R}$ & $W/J=5$ & $W/J=10$ & $W/J=16.5$ & $W/J=20$ \\ \hline
$L=10$       & -       & -        & 500        & -        \\ \hline
$L=12$       & 125     & 500      & 500        & 1000     \\ \hline
$L=16$       & 100     & 400      & 400        & 800      \\ \hline
$L=20$       & 50      & 200      & 200        & 400      \\ \hline
$L=24$       & 25      & 100      & 100        & 200      \\ \hline
$L=28$       & 25      & 50       & 200        & 200      \\ \hline
$L=32$       & 20      & 50       & 50         & 100      \\ \hline
$L=36$       & 20      & 50       & 50         & 50       \\ \hline
$L=40$       & -       & -        & 50         & -        \\ \hline
\end{tabular}
\caption{Number of Hamiltonian realizations $N_\text{R}$ in the 3D Anderson model for a given system size $L$ and disorder strength $W/J$. A dash indicates that we did not generate data for the respective parameters.}
\label{tab:3D_And_N_R}
\end{table}

Our results for the 3D Anderson model are averaged over $N_\text{R}$ different Hamiltonian realizations.
In general, larger values of $N_{\rm R}$ are necessary for larger disorder strengths $W/J$ and since we observe that the transition probabilities and the imbalance exhibit self-averaging with increasing the system size $L$, we decrease $N_\text{R}$ with increasing $L$. 
Whenever $N_{\rm R}$ is not specified in the figure caption, its value for given parameters $L$ and $W/J$ are listed in Tab.~\ref{tab:3D_And_N_R}.
For the 1D Aubry-André model, however, we do not observe any efficient self-averaging of the quantities under investigation, and hence we always average over $N_\text{R}=50$ different realizations of the global phase $\phi$.

\FloatBarrier
\bibliographystyle{biblev1}
\bibliography{references,references1,references2}

\end{document}